\documentclass{nature}
\usepackage{graphicx,psfrag}
\usepackage{mathrsfs}
\usepackage{amsmath,amsfonts,amssymb}
\usepackage{multirow}
\usepackage{comment}
\usepackage{units}
\usepackage{enumerate}
\usepackage[normalem]{ulem} 
\usepackage{longtable}
\usepackage[rgb]{xcolor}
\usepackage{gensymb}
\usepackage{adjustbox}
\usepackage[none]{hyphenat}
\usepackage{multibib}
\usepackage{caption}
\newcites{New}{References}

\newcommand{\mn}{{Mon. Not. R. Astron. Soc.}}
\newcommand{\mnras}{\mn}
\newcommand{\aj}{{"Astron. J."}}
\newcommand{\apj}{{Astrophys. J.}}
\newcommand{\apjl}{{Astrophys. J. Lett.}}
\newcommand{\apjs}{{Astrophys. J. Supp.}}

\newcommand{\aap}{{Astron. Astrophys.}}
\newcommand{\nat}{{Nature}}
\newcommand{\pasj}{{PASJ}}

\newcommand{\pasp}{{Pub. Ast. Soc. Pac.}}

\newcommand{\araa}{Annual Review of Astronomy and Astrophysics}

\title{A dense $\mathbf{0.1 M_{\rm \odot}}$ star in a 51-minute orbital period eclipsing binary}

\author{Kevin B. Burdge$^{1,2}$ $^{*}$, Kareem El-Badry$^{3,4,5}$, Thomas R. Marsh$^{6}$, Saul Rappaport$^{1,2}$, Warren R. Brown$^{3}$, Ilaria Caiazzo$^{7}$, Deepto Chakrabarty$^{1,2}$, V. S. Dhillon$^{8,9}$, Jim Fuller$^{7}$, Boris T. G\"ansicke$^{6}$, Matthew J. Graham$^{7}$, Erin Kara$^{1,2}$, S.~R. Kulkarni$^{7}$, S. P. Littlefair$^{8}$, Przemek Mr{\'o}z$^{10}$, Pablo Rodr\'\i guez-Gil$^{9,11}$, Jan van Roestel$^{7}$, Robert A. Simcoe$^{1,2}$, Eric C. Bellm$^{12}$, Andrew J. Drake$^{7}$, Richard G. Dekany$^{13}$, Steven L. Groom$^{14}$, Russ R. Laher$^{14}$, Frank J. Masci$^{14}$, Reed Riddle$^{13}$, Roger M. Smith$^{13}$, \& Thomas A. Prince$^{7}$}

\begin{document}

\maketitle

\begin{affiliations}
 \item Department of Physics, Massachusetts Institute of Technology, Cambridge, MA 02139, USA
 \item Kavli Institute for Astrophysics and Space Research, Massachusetts Institute of Technology, Cambridge, MA 02139, USA
 \item Center for Astrophysics, Harvard \& Smithsonian, 60 Garden Street, Cambridge, MA 02138
 \item Harvard Society of Fellows, 78 Mount Auburn Street, Cambridge, MA 02138
 \item Max-Planck Institute for Astronomy, K{\"o}nighstuhl 17, D-69117 Heidelberg, Germany
 \item Department of Physics, University of Warwick, Coventry CV4 7AL, UK
  \item Division of Physics, Mathematics and Astronomy, California Institute of Technology, Pasadena, CA, USA
 \item Department of Physics \& Astronomy, University of Sheffield, Sheffield S3 7RH, UK
 \item Instituto de Astrof\'\i sica de Canarias, V\'\i a L\'actea s/n, La Laguna, E-38205 Tenerife, Spain
 \item Astronomical Observatory, University of Warsaw, Al. Ujazdowskie 4, 00-478 Warszawa, Poland
 \item Departamento de Astrof\'\i sica, Universidad de La Laguna, E-38206 La Laguna, Tenerife, Spain
  \item DIRAC Institute, Department of Astronomy, University of Washington, 3910 15th Avenue NE, Seattle, WA 98195, USA
  \item Caltech Optical Observatories, California Institute of Technology, Pasadena, CA 91125, USA
  \item IPAC, California Institute of Technology, Pasadena, CA, USA
\end{affiliations}

\begin{abstract}

In over a thousand known cataclysmic variables (CVs), where a white dwarf is accreting from a hydrogen-rich star, only a dozen have orbital periods below 75 minutes\cite{Green2020,Kato2009,Littlefield2013,Ramsay2014,Augusteijn1996,Kato2014,Thorstensen2002,Breedt2012,Kato2015}. One way to achieve these short periods requires the donor star to have undergone substantial nuclear evolution prior to interacting with the white dwarf\cite{Sienkiewicz1984,Rappaport1984,Nelson1986,Kalomeni2016,Podsiadlowski2003}, and it is expected that these objects will transition to helium accretion. These transitional CVs have been proposed as progenitors of helium CVs\cite{Kalomeni2016,Podsiadlowski2003,Nelemans2010,Goliasch2015,Ramsay2018,El-Badry2021}. However, no known transitional CV is expected to reach an orbital period short enough to account for most of the helium CV population, leaving the role of this evolutionary pathway unclear. Here we report observations of ZTF J1813+4251, a 51-minute orbital period, fully eclipsing binary system consisting of a star with a temperature comparable to that of the Sun but a density 100 times greater due to its helium-rich composition, accreting onto a white dwarf. Phase-resolved spectra, multi-band light curves and the broadband spectral energy distribution allow us to obtain precise and robust constraints on the masses, radii and temperatures of both components. Evolutionary modeling shows that ZTF J1813+4251 is destined to become a helium CV binary, reaching an orbital period under 20 minutes, rendering ZTF J1813+4251 a previously missing link between helium CV binaries and hydrogen-rich CVs.

\end{abstract}

\begin{figure}
\includegraphics[width=6.5in]{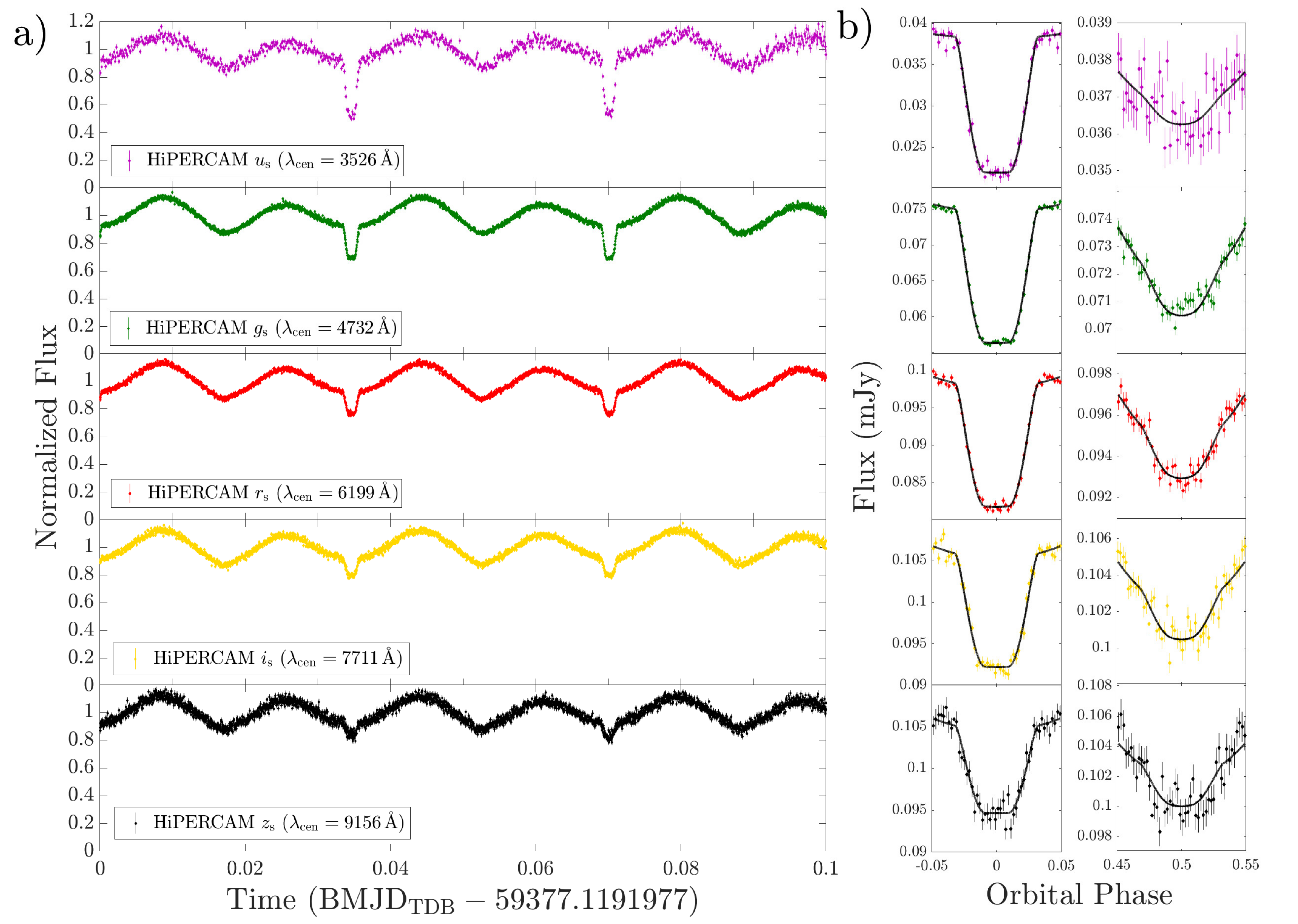}
\linespread{1.3}\selectfont{}
\caption{\textbf{a)} The five color HiPERCAM light curve of ZTF J1813+4251, with filters $u_{\mathrm{s}}$, $g_{\mathrm{s}}$, $r_{\mathrm{s}}$, $i_{\mathrm{s}}$ and $z_{\mathrm{s}}$ arranged from top to bottom. The object exhibits a strong sinusoidal component at twice the orbital frequency due to the tidal deformation of the donor star. The system undergoes a full eclipse, in which the donor fully occults the accreting white dwarf. The depth of the eclipse varies dramatically with wavelength, with the white dwarf contributing half the luminosity in the $u_{\mathrm{s}}$ filter, while only about ten percent in the $z_{\mathrm{s}}$ filter, primarily because the white dwarf preferentially emits its radiation at shorter wavelengths than the cooler donor. \textbf{b)} Best fit light curve models of the primary and secondary eclipses of ZTF~J1813+4251. These models, combined with the spectroscopically derived radial velocity semi-amplitude, allow us to robustly constrain the properties of both components in the system.}

\label{fig:LC}
\end{figure}

Using data from the Zwicky Transient Facility (ZTF)\cite{Bellm2019}, we recently conducted a systematic period search of 1,220,038,476 unique sources using a graphics processing unit (GPU) based implementation of the generalised Lomb-Scargle algorithm\cite{Zechmeister2009}, as part of an ongoing campaign to identify and characterize short period astronomical variables\cite{Burdge2020a} (Methods). This effort led to the identification of ZTF J1813+4251, which exhibited strong ellipsoidal variations due to the tidal deformation of a star\cite{Burdge2019b} and an eclipse, as seen in the lightcurves shown in Figure \ref{fig:LC}. The orbital period of this source is just 51~minutes, despite a spectral energy distribution indicating that it hosts a $~6000\rm\,K$ star of spectral type F. The \emph{Gaia} eDR3 astrometric solution suggests that the source is nearly a kiloparsec away, making it  significantly less luminous than an F type main-sequence star given its apparent brightness. This ultracompact binary system had been excluded from previous searches for ultracompact binaries\cite{Burdge2020a} because of its red color relative to other ultracompact systems.

\begin{figure}
\includegraphics[width=6.5in]{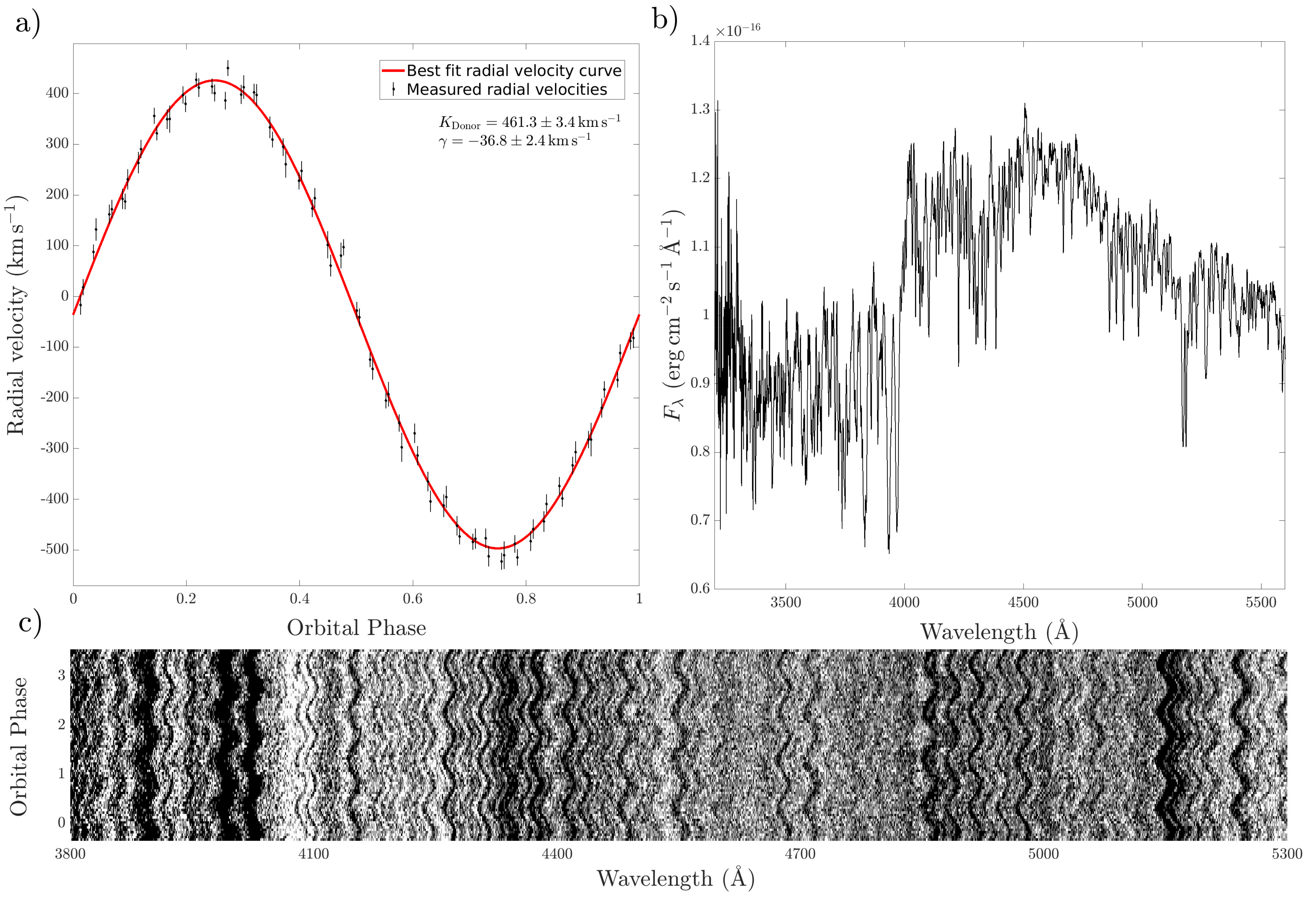}
\linespread{1.3}\selectfont{}
\caption{\textbf{a)} A sinusoidal fit to the measured radial velocities of the donor star in ZTF J1813+4251, with a best-fit velocity semi-amplitude of $K_2=461.3\pm3.4\rm \, km\, s^{-1}$ and systemic velocity of $\gamma=-36.8\pm2.4\rm \, km\, s^{-1}$. \textbf{b)} The blue side of the Keck LRIS spectrum, coadded in the rest frame of the donor star. The spectrum exhibits a large number of narrow metal lines, including those from calcium, iron and magnesium, characteristic of main sequence stars with spectral types similar to the Sun. \textbf{c)} The trailed blue Keck LRIS spectra, illustrating the significant wavelength shifts in the spectrum caused by the large Doppler shifts of the donor. The trailed spectra reveal these Doppler shifts occur in the large number of narrow lines associated with the donor spectrum.}

\label{fig:Spectrum}
\end{figure}

\begin{figure}
\includegraphics[width=6.5in]{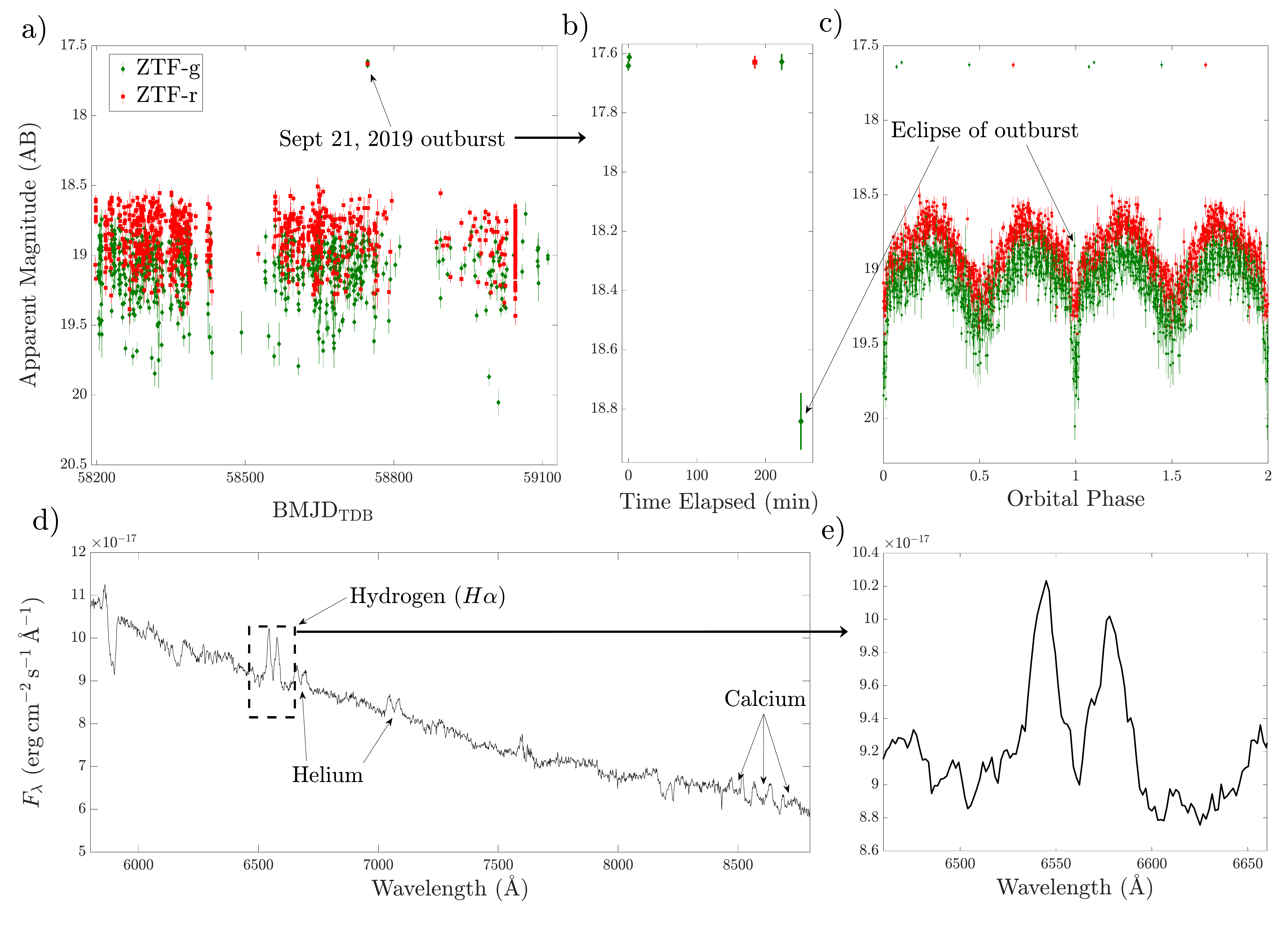}
\linespread{1.3}\selectfont{}
\caption{\textbf{a)} The ZTF light curve of ZTF J1813+4251. The system has been in quiescence in all ZTF observations, except for those taken on September 21, 2019, in which the system brightened by over a factor of two. \textbf{b)} A zoom in of the ZTF data taken on September 21, 2019, during the outburst. The final of the four measurements during the outburst is significantly fainter because it occurred during the primary eclipse of the system. \textbf{c)} The phase-folded ZTF data, illustrating that the faint point which occurred during the outburst coincided with the primary eclipse in the system. \textbf{d)} The red side of the Keck LRIS spectrum, coadded in the rest frame of the accreting white dwarf. Double-peaked emission lines associated with the blueshifted and redshifted components of an accretion disk are clearly visible. This disk exhibits emission associated with several elements, including hydrogen, helium and calcium. \textbf{e)} The profile of the hydrogen H$\alpha$ emission feature, the strongest disk feature in the optical spectrum.}

\label{fig:Red_Spectrum}
\end{figure}

On June 4 and July 5 2021, we obtained phase-resolved spectroscopic observations of ZTF J1813+4251 using the Low Resolution Imaging Spectrometer (LRIS)\cite{Oke1995} on the 10-m W.\ M.\ Keck I Telescope on Mauna Kea (Methods). Figure \ref{fig:Spectrum} illustrates these observations, revealing a late F type stellar spectrum blanketed with metallic absorption lines characteristic of main sequence stars. These absorption lines Doppler shift with a large velocity semi-amplitude of $461.3\pm3.4\rm \, km\, s^{-1}$ due to the short orbital period of the binary. Additionally, double-peaked emission lines of hydrogen, helium, and calcium reveal an accretion disk, most prominently visible in the red half of the spectrum, as seen in Figure \ref{fig:Red_Spectrum}.

On June 12, 2021, we obtained high speed images of ZTF J1813+4251 with the quintuple-beam high speed photometer HiPERCAM\cite{Dhillon2021} on the 10.4-m Gran Telescopio Canarias (GTC) on La Palma. This multi-color high signal-to-noise (SNR) light curve, illustrated in Figure \ref{fig:LC}, revealed that the depth of the eclipse of the white dwarf varies as a function of wavelength, with a depth of $\sim 50$ percent in the HiPERCAM-$u_{\mathrm{s}}$ filter centered at $3526\,\mathrm{\AA}$, but only a $\sim 10$ percent depth in the HiPERCAM-$z_{\mathrm{s}}$ filter centered at $9156\,\mathrm{\AA}$. This dependence of the eclipse depth on wavelength indicates that the accreting white dwarf contributes a larger share of the luminosity at short wavelengths, due to it having a substantially higher surface temperature ($T_{\mathrm{WD}}=12600\pm500\,\rm K$) than the donor star ($T_{\mathrm{Donor}}=6000\pm90\,\rm K$). There are two photometric maxima per orbit due to ellipsoidal variations, and the high signal-to-noise (SNR) HiPERCAM light curve also revealed a pronounced ``O'Connell'' effect\cite{O'Connell1951}, in which the peak flux of these alternating maxima is substantially different due to a modulation component at the orbital frequency. This effect may be due to a spot on the donor star, which is tidally locked, and thus completes a rotation every orbital period.

The accretion disk seen in panels (d) and (e) of Figure \ref{fig:Red_Spectrum} indicates that the donor star in ZTF J1813+4251 is transferring matter to the white dwarf, and thus is filling its Roche-lobe. The Roche-lobe has a scale-invariant geometry dependent only on the mass ratio of the system, and thus the geometry of the donor depends only on this mass ratio\cite{Paczyinski1971}. In a binary system undergoing a total eclipse, the time between mid-ingress and mid-egress depends only on the geometry of the donor and the inclination of the system. This means that the eclipses in ZTF J1813+4251's light curve constrain the system to a unique mass ratio vs inclination relation\cite{Chanan1976}. Additionally, the depths of the primary and secondary eclipse constrain a unique radius ratio between the white dwarf and that of the donor, which, when combined with the overall geometry of the primary eclipse, yields a unique solution for the radii divided by the semi-major axis (the scaled radii), orbital inclination, and thus mass ratio, because the system is Roche lobe filling. These constraints, together with the precise donor radial velocity semi-amplitude measured from the spectra (see the coadded spectrum in Figure \ref{fig:Spectrum}), allow for a robust determination of the system parameters on the basis of only Roche lobe geometry and Kepler's laws (Methods). We report these system parameters in Table 1. Because the donors in transitional CVs are still transferring significant hydrogen and have not switched to primarily helium accretion yet, as seen in a helium CV (also known as AM CVns), these objects often exhibit a mixture of hydrogen disk lines along with unusually strong helium lines in their spectra as a result of their slowly transitioning to transferring mainly helium, and as result, we see in panel (d) of Figure \ref{fig:Red_Spectrum} that the disk also exhibits significant helium emission.

A Niels Gehrels Swift observatory (Swift) X-ray telescope (XRT) observation of the system revealed no detectable X-ray flux, with the X-ray luminosity of the system, $L_{\mathrm{x}}$, constrained to a $3\sigma$ upper limit of $L_{\mathrm{x}}<1.22\times 10^{31}\,\rm erg \, s^{-1}$ given an assumed distance of $891^{+174}_{-135}\,\mathrm{pc}$ inferred from \emph{Gaia} EDR3\cite{Bailer-Jones2021}, an upper limit consistent with the typical x-ray luminosity of a non-magnetic cataclysmic variable\cite{Mukai2017} at these energies. The source was detected with the Swift Ultraviolet Optical Telescope in the ultraviolet, constraining the temperature of the accreting white dwarf (Methods).

As seen in Figure \ref{fig:Red_Spectrum}, archival ZTF data captured an outburst of the object on September 21, 2019, brightening by over a factor of two relative to its quiescent brightness in the ZTF-$g$ and ZTF-$r$ filters. The final observation of the night during which the outburst was detected took place during the totality of the primary eclipse, and as a result the flux was significantly attenuated, indicating that this brightening must have originated from the accretor. Because the white dwarf was completely eclipsed during this phase, but the flux was still partially elevated relative to quiescence, we can infer that the luminosity must be originating from the accretion disk, which was not completely occulted at this orbital phase.

We used Modules for Experiments in Stellar Astrophysics\cite{Paxton2011} (MESA) to model the past and future evolution of systems like ZTF J1813+4251, and Figure \ref{fig:Evolution} illustrates these evolutionary tracks. Currently, ZTF J1813+4251 exhibits an elevated temperature because the white dwarf is rapidly removing matter from the donor as the orbital period shrinks, causing its radius to decrease at roughly constant luminosity. We expect to be able to detect this orbital period evolution over the next decade, as using the HiPERCAM light curve we are able to measure the mid-eclipse time, $T_{0}$, to a precision of $\approx0.16\,\rm s$, which will allow us to test whether the angular momentum loss in the binary deviates from that expected purely due to general relativity. Assuming that the system is evolving purely according to general relativistic orbital decay, the eclipse time will deviate from that expected if the orbital period were constant by about $7\, \rm s$ over the next ten years\cite{Burdge2019a}. However, evolutionary models (Methods) indicate that magnetic braking may still be responsible for as much as two thirds of the orbital angular momentum loss in the system, and there are other effects which may influence the orbital evolution as well such as mass-transfer. As seen in panels (a) and (d) of Figure \ref{fig:Evolution}, when the system evolves to shorter orbital periods, the last remnants of hydrogen will be removed from the donor, eliminating its fuel source, and its helium rich remnant will begin to transition to being supported by electron degeneracy pressure. At this stage, which our models predict will occur in approximately 75 million years, ZTF J1813+4251 will have completed its transition to a helium CV system, reaching a period minimum around 18 minutes. Due to it being supported by electron degeneracy pressure, ZTF J1813+4251's structure will undergo an important transition, in which it will begin to expand in response to mass loss, rather than shrink. This will cause further accretion to decrease the donor's density as the system evolves to longer orbital periods, and the donor will rapidly cool as a result of adiabatic expansion in response to mass loss. We estimate that after the period minimum, the system will spend three hundred million years evolving back out to an orbital period of about 30 minutes as a helium CV, thus spending a much larger fraction of its remaining life in this state than in its current short-lived orbital decay phase as a transitional CV.

The discovery of ZTF J1813+4251 demonstrates that evolved CV donors can reach orbital periods below 20 minutes, and likely form helium CVs. Other known examples of candidate transitional CVs\cite{Kato2009,Littlefield2013,Ramsay2014,Augusteijn1996,Thorstensen2002,Breedt2012,Kato2015} below the period minimum are either dominated entirely by the flux of an accretion disk, making it difficult to characterize the properties of the donor star, or exhibit a detectable donor which is not evolved enough to reach orbital periods short enough to account for the full range of periods seen in the helium CV population. Additionally, none of the known transitional CVs exhibit a clean radial velocity semi-amplitude paired with a detectable full primary and secondary eclipse like that seen in ZTF J1813+4251, making ZTF J1813+4251 the only transitional CV with robustly characterized system parameters. As seen in panel A of Figure 4, the most important property in predicting how evolved the donor is -- and thus what period minimum it can reach -- is its effective temperature. As seen in panel A of figure \ref{fig:Evolution}, we also modelled EI~Psc\cite{Thorstensen2002}, which is one of the few transitional CVs with enough parameter estimates to construct such models. Like ZTF J1813+4251, we expect this system to evolve into a helium CV, however it reaches a period minimum of around 30 minutes rather than below 20 minutes. ZTF J1813+4251, which has a shorter orbital period and significantly hotter donor than any currently known transitional CV candidate below the period minimum, demonstrates that the donors of these systems can reach much more extreme states than seen in previously identified systems, and solidifies the role of this evolutionary channel in the formation of helium CVs, including those with orbital periods short enough to emit significantly in the LISA gravitational wave band. ZTF J1813+4251, with its well-characterized components, will serve as an anchor point for models of binary evolution connecting the population of hydrogen accreting CVs with evolved donors at longer orbital periods and the population of ultracompact helium CVs. 

\begin{figure}
\includegraphics[width=6.5in]{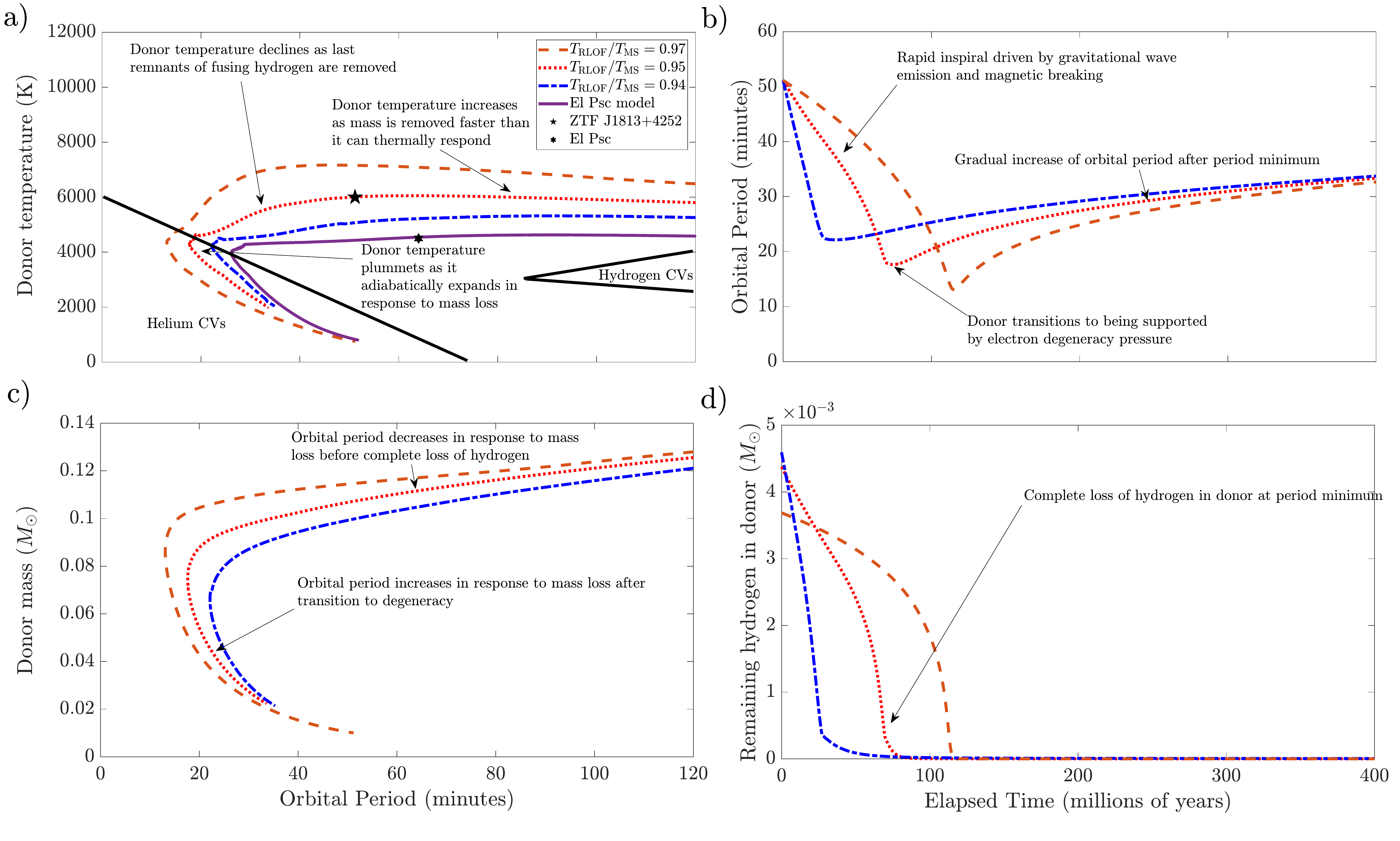}
\linespread{1.3}\selectfont{}
\caption{ \textbf{Evolutionary tracks of ZTF J1813+4251 generated using Modules for Experiments in Stellar Astrophysics (MESA).} \textbf{a)} The donor's temperature evolution as a function of orbital period in transitional CVs like ZTF J1813+4251. The wide dashed orange line, dotted red line and alternating dashed blue line indicate MESA evolutionary sequences corresponding to a transitional CV where mass loss from the donor commenced at 97, 95 and 94 percent of its main sequence lifetime, and the black star indicates the position of ZTF J1813+4252 on the red dotted track. These sequences host warmer donors than normal hydrogen CVs, whose location are indicated on the right of the plot, and evolve into the region occupied by helium CVs in the lower left corner. For comparison, we also illustrate a track for El Psc\cite{Thorstensen2002}, another transitional CV candidate which will likely reach a period minimum of just under 30 minutes. \textbf{b)} The future orbital period evolution of the MESA tracks. The track corresponding to ZTF J1813+4251 will reach a period minimum of about 18 minutes in approximately 75 million years, and will spend the next 300 million years evolving back out to a period of about half an hour as a helium CV. \textbf{c)} The evolution of the donor mass as a function of orbital period, reaching just a few hundredths of a solar mass as the tracks evolve out to longer orbital periods as helium CVs. \textbf{d)} The hydrogen remaining in the donor over the course of its future evolution. At around the period minimum, the donor star loses all remaining hydrogen. (See also Kalomeni et al. (2016)\cite{Kalomeni2016} for a comprehensive discussion of the evolution of these systems.)}

\label{fig:Evolution}
\end{figure}

\begin{table}
\renewcommand{\thetable}{\arabic{table}}
\centering
 \label{tab:Parameters}
\begin{tabular}{cc}

\hline
\hline
$\rm Right\,\,ascension$ &$18^\mathrm{h}\,\,13^\mathrm{m}\,\,11.13^\mathrm{s}$ \\
   \hline
   $\rm Declination$ &$+42^\circ\,\,51'\,\,50.4''$ \\
   \hline
   $\rm Proper\,\,motion\,\,in\,\,right\,\,ascension$ &$-12.32\pm0.19\,\mathrm{mas\,yr^{-1}}$\\
   \hline
   $\rm Proper\,\,motion\,\,in\,\,declination$ &$-2.66\pm0.19\,\mathrm{mas\,yr^{-1}}$\\
   \hline
   $\mathrm{Parallax}\,\,(\sigma)$ &$1.20\pm0.16\,\mathrm{mas}$\\
   \hline
   $\mathrm{Distance}\,\,(D)$ &$891^{+174}_{-135}\,\mathrm{pc}$\\
   \hline
   $\mathrm{Systemic\,\,velocity}\,\,(\gamma)$ &$-36.8\pm2.4\,\mathrm{km\,s^{-1}}$\\   
   \hline
   \hline
   $\mathrm{Orbital\,\,period}\,\,(P_{\mathrm{b}})$ & $3069.64398\pm-0.00015\,  \mathrm{s}$  \\
   \hline
    $\mathrm{Orbital\,\,period\,\,derivative}\,\,(\dot{P_{\mathrm{b}}})$ & $-0.29\pm1.15 \times 10^{-12}\,  \mathrm{s}\,\mathrm{s}^{-1}$  \\
   \hline
   $\mathrm{Time\,\,of\,\,superior\,\,conjunction}\,\,(T_{0})$ & $59377.1538184\pm0.0000020\,   \mathrm{BMJD_{TDB}}$  \\
   \hline
   $\mathrm{Radial\,\,velocity\,\,of\,\,donor}\,\,(K_{\mathrm{Donor}})$ & $461.3\pm3.4\,\mathrm{km\,s^{-1}}$  \\
   \hline
   $\mathrm{Orbital\,\,inclination}\,\,(i)$ & $78.80\pm0.18\,\mathrm{deg}$  \\
    \hline
   $\mathrm{Semi\,\,major\,\,axis}\,\,(a)$ & $0.4000\pm0.0041\,R_{\rm \odot}$  \\
   \hline
   \hline
   $\mathrm{White\,\,dwarf\,\,mass}\,\,(M_{\mathrm WD})$ & $0.562\pm0.015\,M_{\rm \odot}$  \\
   \hline
   $\mathrm{Donor\,\,mass}\,\,(M_{\mathrm{Donor}})$ & $0.1185\pm0.0067\,M_{\rm \odot}$  \\
   \hline
   $\mathrm{White\,\,dwarf\,\,radius}\,\,(R_{\mathrm WD})$ & $0.01374\pm0.00023\,R_{\rm \odot}$  \\
   \hline
   $\mathrm{Donor\,\,radius}\,\,(R_{\mathrm{Donor}})$ & $0.1017\pm0.0019\,R_{\rm \odot}$  \\
   \hline
   $\mathrm{White\,\,dwarf\,\,temperature}\,\,(T_{\mathrm WD})$ & $12600\pm500\,\rm K$  \\
   \hline
   $\mathrm{Donor\,\,temperature}\,\,(T_{\mathrm{Donor}})$ & $6000\pm80\,\rm K$  \\
   \hline
   \hline

\end{tabular}
\caption{Table of parameters (The first five parameters are the astrometric solution reported by \emph{Gaia} eDR3\cite{Gaia2021}, at Epoch J2016.0 and Equinox J2000.0).}
\end{table}

\begin{addendum}
 \item K.B.B. is a Pappalardo Postdoctoral Fellow in Physics at MIT and thanks the Pappalardo fellowship program for supporting his research.
 The design and construction of HiPERCAM was funded by the European Research Council under the European Union’s Seventh Framework Programme (FP/2007-2013) under ERC-2013-ADG Grant Agreement no. 340040 (HiPERCAM). VSD and HiPERCAM operations are supported by STFC grant ST/V000853/1. 
 TRM and BTG acknowledge support from the UK's Science and Technology Facilities Council (STFC), grant ST/T000406/1.
This project has received funding from the European Research Council (ERC) under the European Union’s Horizon 2020 research and innovation programme (Grant agreement No. 101020057). 

Based on observations made with the Gran Telescopio Canarias (GTC), installed at the Spanish Observatorio del Roque de los Muchachos of the Instituto de Astrofísica de Canarias, on the island of La Palma.

 Based on observations obtained with the Samuel Oschin Telescope 48-inch Telescope at the Palomar Observatory as part of the Zwicky Transient Facility project. ZTF is supported by the National Science Foundation under Grant No. AST-1440341 and a collaboration including Caltech, IPAC, the Weizmann Institute for Science, the Oskar Klein Center at Stockholm University, the University of Maryland, the University of Washington, Deutsches Elektronen-Synchrotron and Humboldt University, Los Alamos National Laboratories, the TANGO Consortium of Taiwan, the University of Wisconsin at Milwaukee, and Lawrence Berkeley National Laboratories. Operations are conducted by COO, IPAC, and UW.
 
Some of the data presented herein were obtained at the W.M. Keck Observatory, which is operated as a scientific partnership among the California Institute of Technology, the University of California and the National Aeronautics and Space Administration. The Observatory was made possible by the generous financial support of the W.M. Keck Foundation. The authors wish to recognize and acknowledge the very significant cultural role and reverence that the summit of Mauna Kea has always had within the indigenous Hawaiian community. We are most fortunate to have the opportunity to conduct observations from this mountain.
 \item[Competing Interests] The authors declare that they have no
competing financial interests.
 \item[Correspondence] Correspondence and requests for materials
should be addressed to K.B.B.~(email: kburdge@mit.edu).
\end{addendum}

\newpage

\begin{methods}

\subsection{Period Search}

We performed a period search using a GPU implementation of the generalised Lomb-Scargle algorithm\citeNew{Zechmeister2009}, an analog of the Fourier transform optimized for data with non-equispaced sampling. This search was based on newly generated forced point spread function (PSF) photometry performed at the coordinates of all Pan-STARRS1 sources in ZTF images over a period range of $250\rm \, days$ down to just $2\rm \, minutes$. By systematically searching all 1,220,038,476 unique sources in the Pan-STARRS1 source catalog\cite{Chambers2016} with more than 50 ZTF epochs, we eliminated biases of previous searches relying on colour or astrometric based selections\cite{Burdge2020a}, which represents a major advancement in probing for short period astrophysical variables across the northern sky. We searched a total of 1,461,592 trial frequencies per source, corresponding to an oversampling factor of 2. In total, given that we searched 1,220,038,476 sources, this means a total of $1.78\times 10^{15}$ trial frequencies were searched, and each source had on average approximately $\sim 10^3$ epochs. This search was completed in six weeks on a single desktop containing four Nvidia 2080 Ti GPUs, and this same desktop was used to perform the forced PSF photometry of all Pan-STARRs sources in ZTF images.

\subsection{Keck LRIS observation}

We obtained the Keck LRIS observations using the 600/4000 grism as the dispersive element on the blue arm and the 600/7500 grating on the red arm. We used $2\times 2$ binning on both channels to reduce both readout noise and readout time, and used an exposure time of $120\,\rm s$ to mitigate the impact of orbital Doppler smearing on individual exposures. We reduced the Keck LRIS observations using a publicly available LPIPE automated data reduction pipeline\cite{Perley2019}.

\subsection{Keck ESI observation}
We obtained sixteen $120\,\rm s$ medium-resolution spectra using the Echellette Spectrograph and Imager (ESI) on the 10-m W.\ M.\ Keck II Telescope on Mauna Kea. These exposures were obtained in the cross-dispersed echelle mode and were binned $2\times 2$ in order to reduce readout overheads and readout noise. The data were reduced using the MAuna Kea Echelle Extraction (MAKEE) pipeline.

\subsection{HiPERCAM observation}

We reduced the HiPERCAM data using the publicly available pipeline\cite{Dhillon2021}, and selected several comparison stars in each filter to ensure accurate absolute flux calibrations. The pipeline performed aperture photometry with a dynamic full-width-half-maximum. We used $9, 3, 3, 3 $ and $3\,\rm s$ exposures in $u_{\mathrm{s}}$, $g_{\mathrm{s}}$, $r_{\mathrm{s}}$, $i_{\mathrm{s}}$, and $z_{\mathrm{s}}$, respectively. The charge coupled devices (CCDs) were operated with conventional amplifiers using frame-transfer to effectively eliminate readout overheads. The total observation lasted 150~minutes, or about three orbital cycles.

\subsection{Swift XRT observation}

We obtained a 2,777 s Swift observation of ZTF J1813+4251 (observation ID 00014355002) in order to constrain the X-ray flux associated with the mass transfer in the system. We do not detect the source with significance, with just two counts within an 18 arcsecond aperture, corresponding to a $3\sigma$ upper limit of $0.0039\,\rm cnts\,s^{-1}$ at $0.2-10\,\rm keV$. We used WebPIMMs to estimate the source flux. Assuming a $3\,\rm keV$ thermal Bremsstrahlung model characteristic of non-magnetic CVs, and a column density of $N_H=5.58\times 10^{19} \,\rm{cm}^{-2}$ based on the estimated reddening of $E(g-r)=0.01$, we place a $3\sigma$ upper limit on the $0.2-10\,\rm keV$ X-ray luminosity of $L_{\mathrm{x}}<1.23\times10^{31}\,\rm erg\,s^{-1}$. We did not use the NASA HEASARC WebPIMMs column density estimator tool, as this estimates the column density at infinity, and thus is a significant overestimate for a nearby source like ZTF J1813+5242, and instead we based the column density estimate on more detailed dust maps which take into account the distance to the source\cite{Green2019}.

\subsection{Swift UVOT observation}

The 2,777\,s Swift observation (observation ID 00014355002) also included data obtained with the UVOT instrument in the UVW2 filter. We computed an average apparent magnitude of the source using the uvotsource tool and a 5 arcsecond aperture centered on the source. Due to a nearby bright star, the background near the source was significantly elevated, and thus we used an annulus background aperture centered on the source, with an inner radius of 7.5 arcsecond and an outer radius of 15 arcsecond. We find an apparent magnitude of $21.20\pm0.1\mathrm{\,(statistical)}\pm0.03\mathrm{\,(systematic)}\,m_{\rm AB}$.

\subsection{Spectral energy distribution (SED) analysis}
We estimated the spectroscopic properties of the donor in ZTF J1813+4251 by fitting atmospheric models to the in-eclipse spectral energy distribution, as our spectra are composed of contributions from both the white dwarf and the accretion disk with exposures too long to have obtained a completely in-eclipse spectrum. We use the in-eclipse apparent magnitudes of the system to obtain an uncontaminated SED of the donor star, and fit the BT-NextGen (GNS93) atmospheric models for low mass stars\cite{Allard2012} to the SED, allowing the metalicity, flux scale and temperature to vary, and fixing the surface gravity to $\log(g)_{\mathrm Donor}=5.43\,\rm (cgs)$ based on an estimate derived from an initial light curve model which used the \emph{Gaia} eDR3 estimated temperature of $6088\,\rm K$ for the donor. Using the UltraNest kernel density estimator\cite{Buchner2014,Buchner2017,Buchner2021}, we sampled over these parameters and obtained a donor temperature estimate of $T_{\mathrm{Donor}}=6000\pm80\,\rm K$. The metallicity preferred the lowest possible value covered by the atmospheric model grids, $Z_{\mathrm{Donor}}=-1.5\,\rm [Fe/H]$. Because of the evolved nature of the donor, we consider this metallicity estimate based on main-sequence models to be dubious, and thus do not report it in the table; we suspect the uncertainties in the $T_{\mathrm{Donor}}$ are also likely underestimated, though we are unable to appropriately quantify the model error associated with using model atmospheres for main-sequence stars on such an extreme evolved star. We overplot the best fit atmospheric models with the donor and accretor SEDs in Extended Data Figure \ref{fig:SED_fit}.

We are also able to precisely measure the contribution of the white dwarf to the SED because of the eclipses in the light curve, and thus we performed an SED fit to it as well in order to estimate $T_{\mathrm WD}$. A fit to the optical SED using a white dwarf model\cite{Tremblay2009,Tremblay2011} with a fixed $\log(g)_\mathrm{WD}=7.90\,\rm (cgs)$ based on an initial light curve model yields an estimated effective temperature of $T_\mathrm{WD}=12600\pm500\,\rm K$. 

\begin{figure}
\includegraphics[width=6.5in]{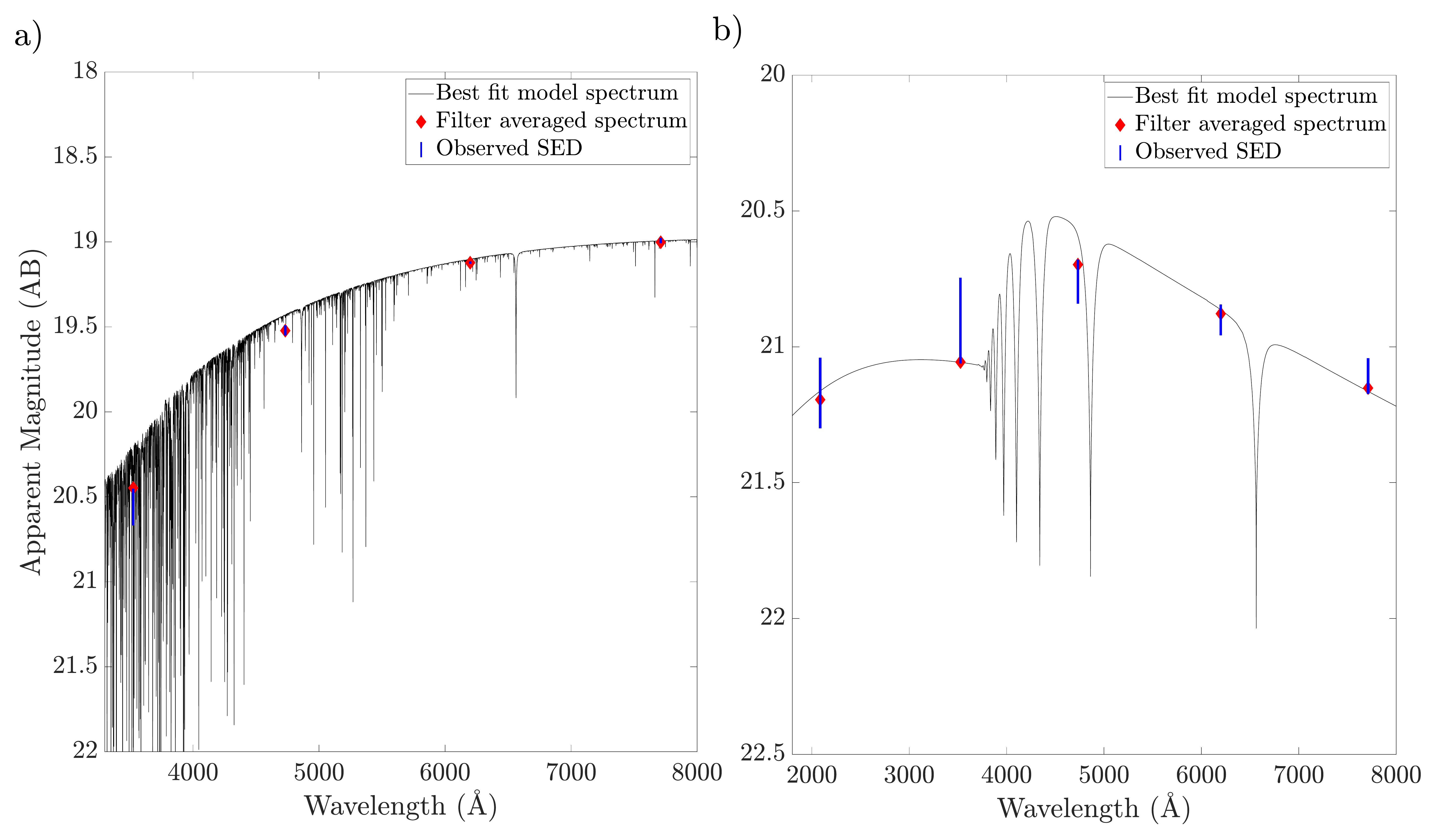}
\renewcommand{\figurename}{Extended Data Figure}
\setcounter{figure}{0}  
\linespread{1.3}\selectfont{}
\caption{\textbf{a)} A fit to the donor star's SED. The blue vertical lines represent the measured donor star SED as inferred from the apparent magnitude during the primary eclipse. Because of the high SNR of the HiPERCAM data, these uncertainties are very small, and we have added in a systematic five percent uncertainty associated with contribution from the accretion disk. The red diamonds represent the filtered averaged apparent magnitudes of the best fit synthetic spectrum, and this synthetic spectrum is plotted in black. \textbf{b)} A fit to the accreting white dwarf SED. The vertical blue lines represent the SED of the white dwarf, as measured by modelling the amount of flux lost during the primary eclipse, when it is occulted by the donor. We added a five percent model error to this SED in the optical to account for contribution from accretion features which might cause the white dwarf to deviate from a standard hydrogen rich DA model spectrum\cite{Tremblay2009,Tremblay2011}. The surface gravity of the white dwarf in the model spectrum was fixed using the values derived from the lightcurve analysis. The red diamonds illustrate the filter averaged apparent magnitudes from the synthetic spectrum, which is plotted in black. In both panels, we have omitted the HiPERCAM $z_s$-band measurement, which exhibits an excess likely associated with the accretion disk. The model deviates from the observed spectrum around the hydrogen absorption lines due to emission lines from the accretion disk around the white dwarf.}

\label{fig:SED_fit}
\end{figure}

\subsection{Atmospheric analysis}

We summed the phase-resolved LRIS spectra in the rest frame of the donor and performed fits with atmospheric models in order to estimate a temperature independently of the SED based analysis. We broaden the spectra by the predicted rotational velocity in the donor and convolve the synthetic spectra with the instrument resolution. We fix the relative contributions of the WD and donor by fixing the radius of each object in our spectroscopic models, as well as the overall distance to the system. Using hydrogen rich white dwarf DA models in combination with the BT-NextGen (GNS93)\cite{Hauschildt1999} low mass main sequence models, we construct a grid of composite spectra to carry out the fitting procedure. Extended Data Figure \ref{fig:Spectrum_fit} shows our best fit, as well as a fit based on the temperatures derived from the SED. We consider both fits acceptable, though the estimated temperatures differ by $\approx 1200 \, \rm K$. We consider the temperature derived from the SED to be more reliable, as it uses the Swift UVOT measurement, which strongly constrains the white dwarf temperature (the estimated spectroscopic temperature of $13780\pm 380\,\rm K$ significantly overestimates the Swift UVOT luminosity).

\begin{figure}
\includegraphics[width=6.5in]{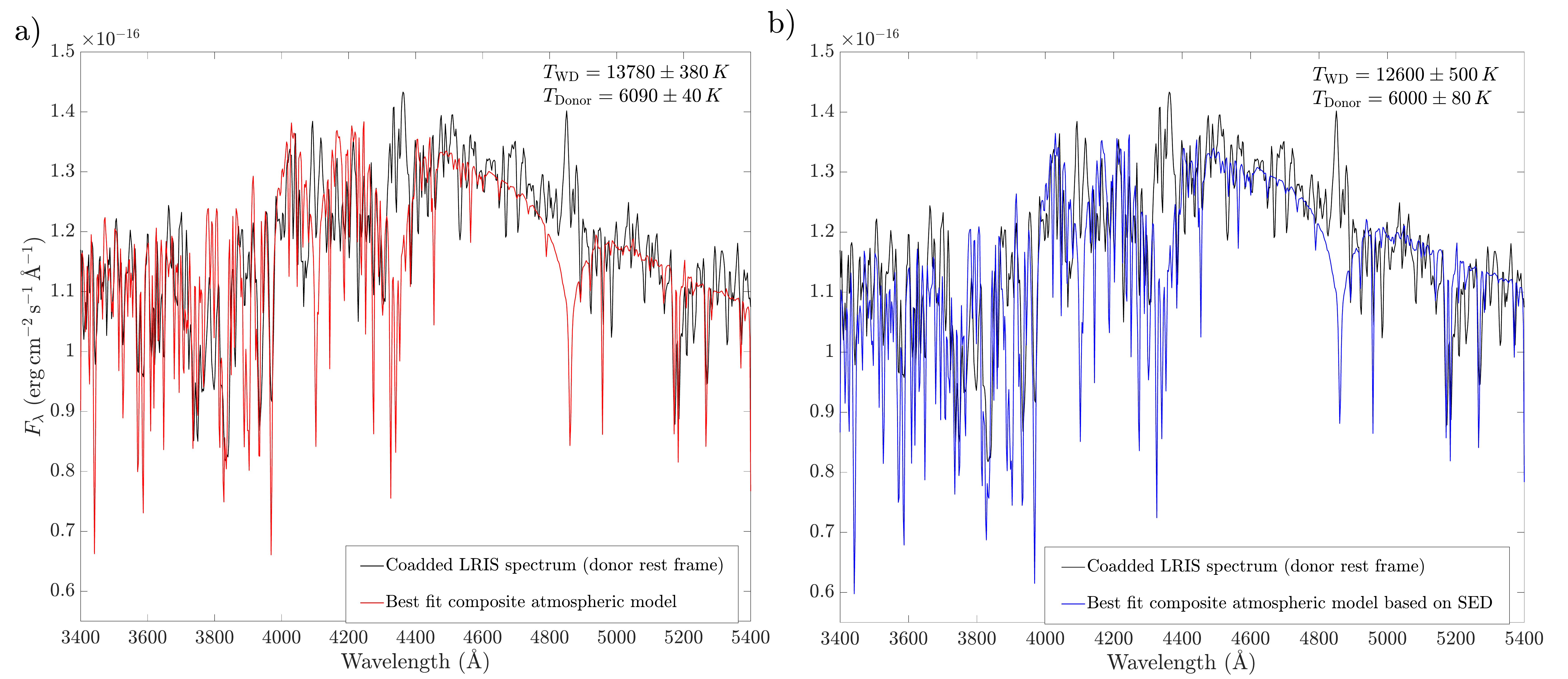}
\renewcommand{\figurename}{Extended Data Figure}
\linespread{1.3}\selectfont{}
\caption{\textbf{a)} A best fit spectral model consisting of a low mass main-sequence model and a hydrogen rich white dwarf model to the observed spectrum of ZTF J1813+4251. The model, shown in red, has been convolved with the resolution of the spectrograph and rotationally broadened, and fit to the spectrum coadded in the rest frame of the donor, shown in black. \textbf{b)} A model spectrum fixed to the parameters derived from the SED analysis. We consider this model more reliable because it takes into account the Swift ultraviolet flux measurement, which strongly constrains the temperature of the white dwarf. The model, shown in blue, slightly underfits the blue part of the spectrum, plotted in black, which may be due to the presence of an accretion disk, or simply a systematic error introduced in the data reduction.}

\label{fig:Spectrum_fit}
\end{figure}

\subsection{Radial velocity analysis}
We measured radial velocities from the Keck LRIS spectra by fitting nine narrow absorption lines with Voigt profiles. We fit the Fe I lines at $4271.7602, 4325.7616, 4383.5447, \\ 4404.7501, 4891.4921, 4920.5028, 4957.5965\,\rm \AA$, the Mg I lines at $4702.9909\,\rm \AA$ and the Ca I line at $4226.73\,\rm \AA$. We forced all lines to share the same Doppler shift and full-width at half-maximum for both the Gaussian and Lorentzian components of the Voigt profile as common free parameters, but allowed the amplitude of each line to vary individually. We used a least squares minimizer to fit each individual spectrum for a Doppler shift relative to the rest frame, and then fit a sinusoidal function to these velocities to derive a semi-amplitude and systemic velocity $\gamma$, as seen in panel a) of Figure \ref{fig:Spectrum}, with uncertainties derived from the covariance matrix of the fit.

We validate the semi-amplitude and systemic velocity with an independent cross-correlation analysis.  We try two cross-correlation spectral templates: a template matched to the spectral type, and a template created from summing the ZTF J1813+4251 spectra shifted to the rest frame of the donor.  We cross-correlate the individual spectra from 4900 to 5600 \AA, a wavelength range filled with narrow absorption lines and dominated by the F-type star.  We find statistically identical results, $K=463.2$ km~s$^{-1}$, consistent within the 1$\sigma$ uncertainty of the published value (Table 1).

\subsection{Rotational broadening analysis}

We coadded the phase-resolved ESI spectra in the rest frame of the donor using the radial velocity semi-amplitude derived from the LRIS spectroscopy. Due to the highly non-linear dispersion in the instrument, only the red end of each echelle order exhibited signal above the read-noise floor, and the SNR of the coadded spectra was too low to obtain a useful constraint on the rotational broadening of the donor. As illustrated in Extended Data Figure \ref{fig:ESI_fit}, we fit four lines, the Fe I lines at $4920.5028, 4957.5965\,\rm \AA$ and the Ca I lines at $6122.22,6162.17\,\rm \AA$. These fits yielded measured rotationally broadened velocities of $159,178,129~{\rm and}~184\,\rm km\,s^{-1}$, with a large scatter due to the low SNR of the spectra, but with values roughly in agreement of the predicted value of $145\,\rm km\,s^{-1}$ (they are on average sligthly higher, but were not corrected for Dopper smearing, which should inflate values by several tens of $\rm km\,s^{-1}$. Future higher SNR moderate to high resolution spectroscopy could yield a more precise measurement, which would serve as a test of the parameters reported in Table 1. 

\begin{figure}
\includegraphics[width=6.5in]{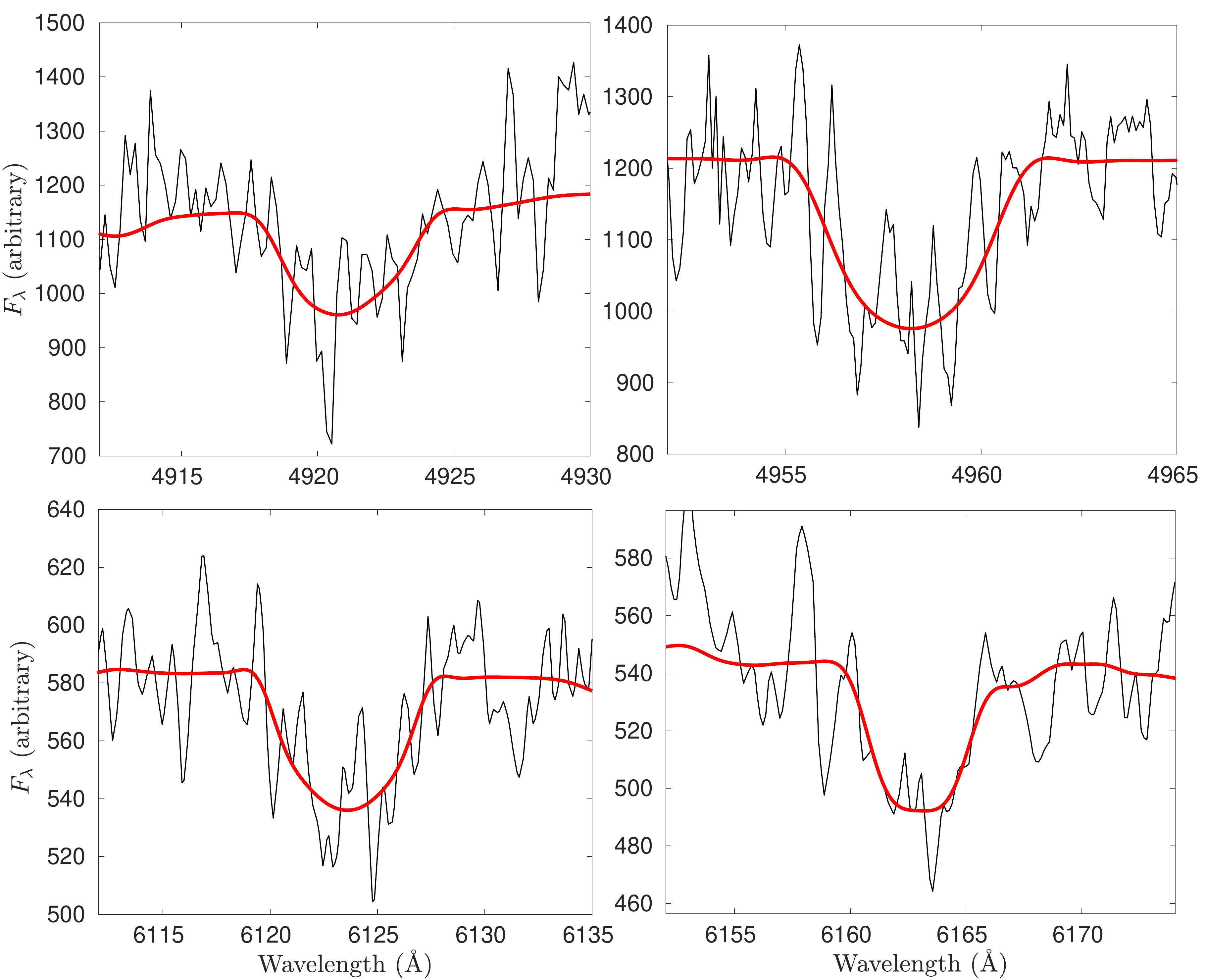}
\linespread{1.3}\selectfont{}
\renewcommand{\figurename}{Extended Data Figure}
\caption{Rotationally broadened atmospheric model fits to the moderate resolution ESI spectra of ZTF J1813+4251. Due to the low SNR of the spectra, we were unable to constrain the rotational broadening of the lines with a precision better than $\sim 50\,\rm km\,s^{-1}$, but the measured values are consistent with the predicted value of  $145\,\rm km\,s^{-1}$.}

\label{fig:ESI_fit}
\end{figure}

\subsection{Light curve analysis}

The core of our analysis was based on the modelling of ZTF J1813+4251's HiPERCAM light curves using the LCURVE light curve modelling code\cite{Copperwheat2010}. LCURVE simulates two stars in a Roche potential, determining the flux at each surface element of the star, taking into account effects such as limb darkening and gravity darkening, and computes the flux as seen from the perspective of an observer at a given inclination and orbital phase. While this code is numerical, below we outline a simplified analytic argument demonstrating why the modelling code is able to robustly constrain the parameters in the system. Extended Data Figure \ref{fig:LC_fit} illustrates this idealized example and provides all relevant analytic expressions.

\begin{figure}
\includegraphics[width=6.5in]{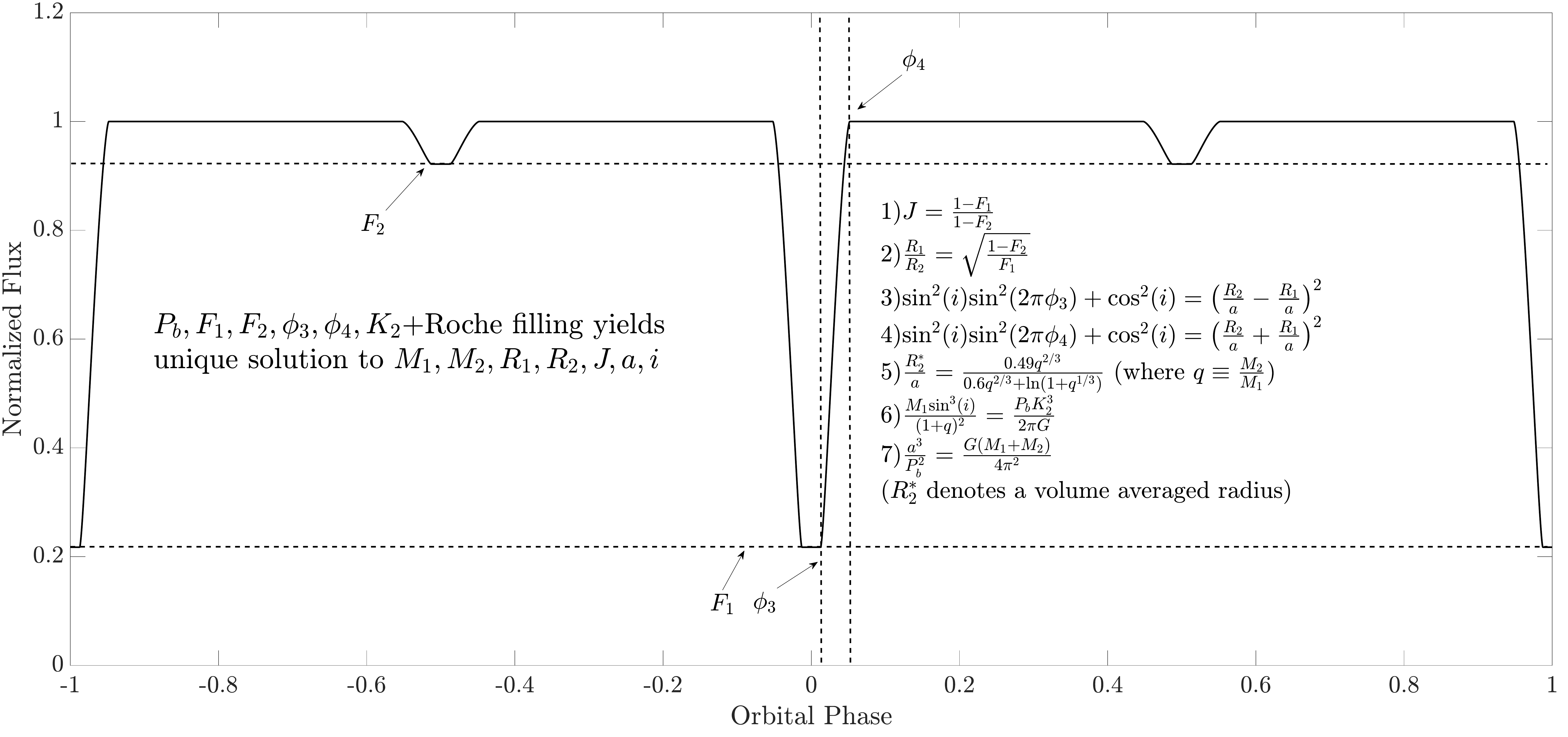}
\renewcommand{\figurename}{Extended Data Figure}
\linespread{1.3}\selectfont{}
\caption{A panel illustrating an idealized version of the basic constraints we obtain by modelling the primary and secondary eclipses of ZTF J1813+5251, which depend only on Roche geometry and Kepler's laws. Using the light curve, we are able to measure the orbital period, $P_\mathrm{b}$, the in eclipse flux levels as a fraction of the out of eclipse flux, $F_1$ and $F_2$, and the third and fourth contact phases of the eclipse, $\phi_3$ and $\phi_4$. By combining these five quantities we can determine from the light curve, with the donor radial velocity semi-amplitude measured from the spectra, $K_2$, and our knowledge that the donor is Roche-lobe-filling, we are able to obtain a robust solution for the two component masses, $M_1$ and $M_2$, the radii of the two components, $R_1$ and $R_2$, the surface brightness ratio, $J$, the semi-major axis of the binary, $a$, and the orbital inclination, $i$. The idealized expressions tying these constraints to the observable quantities are listed in the right half of the figure. We would like to emphasize that these are idealized expressions, and caution readers that there are further important subtleties not discussed here (e.g. in a Roche filling system, the relevant radii in the above equations $R_1$ and $R_2$ are complicated to define because of the ellipsoidal deformation of the components, however, lightcurve modelling codes such as LCURVE take these effects into account--this means for example that the $R_2$ in expressions 3 and 4 is not exactly the same as the $R_2^{*}$ in expression 5, as in the former case, $R_2$ is measured perpendicular to the line between the two stars, whereas in the latter case, $R_2^{*}$ is a volume averaged radius).}

\label{fig:LC_fit}
\end{figure}

Because ZTF J1813+4251 undergoes a total eclipse, in which the white dwarf passes completely behind the donor, we are able to measure three parameters from it: the fractional flux depth in eclipse with respect to the out of eclipse flux, $F_1$, the third contact phase corresponding to the start of egress relative to the mid-eclipse time, $\phi_3$, which measures the duration of the flat bottomed portion of the eclipse when the white dwarf is completely obscured, and the fourth contact phase, $\phi_4$, which measures the end of egress with respect to the mid-eclipse time, and in combination with the third contact phase, describes the duration of egress (one could alternatively use the first and second contact phases, which measure the start and end of ingress, but these are degenerate with the third and fourth contact phases). In an idealized circular orbit involving two spherical objects, the sum and differences of the scaled radii of the two components are constrained by the following two expressions:

\begin{equation}
\sin^2(i) \sin^2(2 \pi \phi_3)+\cos^2(i)=\left(\frac{R_{\mathrm{Donor}}}{a}-\frac{R_{\mathrm{WD}}}{a}\right)^2
\label{eq:radiusratio}
\end{equation}
\begin{equation}
\sin^2(i)\sin^2(2 \pi \phi_4)+\cos^2(i)=\left(\frac{R_{\mathrm{Donor}}}{a}+\frac{R_{\mathrm{WD}}}{a}\right)^2~,
\label{eq:radiusratio2}
\end{equation}where $\frac{R_{\mathrm{WD}}}{a}$, $\frac{R_{\mathrm{Donor}}}{a}$ are the radii of the components, scaled to the semi-major axis. While the primary eclipse gives three parameters to measure, its geometry depends on four physical parameters: the orbital inclination, $i$, the scaled radii of the two components and the surface brightness ratio of the two components, $J$. This means that in order to constrain the parameters of the system from the primary eclipse alone, it is necessary to invoke an additional constraint such as a white dwarf mass-radius relation, or by constraining the surface brightness ratio using temperature estimates of the components, which depend on model atmospheres. However, we detect the secondary eclipse in the system, when the white dwarf transits the donor. This provides one additional geometric constraint, as we can measure the flux depth during the secondary eclipse as a fraction of the out of eclipse flux, $F_2$ (the duration of ingress/egress and the base of the secondary eclipse should be identical to that of the primary, so the only additional parameter is the depth). Using the flux level in the two eclipses, we can constrain the ratio of the radii of the two components, using the following relation:

\begin{equation}
\frac{R_{\mathrm{WD}}}{R_{\mathrm{Donor}}}=\sqrt{\frac{1-F_2}{F_1}}~.
\label{eq:radiusratio3}
\end{equation} 

Additionally, the surface brightness ratio is also constrained by the relative depths of the eclipses, and is given by:

\begin{equation}
J=\frac{1-F_1}{1-F_2}~.
\label{eq:radiusratio3}
\end{equation} 

The above expressions are only applicable for a highly idealized case of two spherical objects transiting each other, and thus only approximate in the case of ZTF~J1813+4251, which hosts a highly distorted donor. The LCURVE modeling simulates the fully distorted donor, and thus accounts for the complexities accompanied by this. 

The fact that the donor is filling its Roche lobe (which we know because the system is mass transferring) allows us to place an additional constraint, as a Roche-lobe-filling donor in a binary which undergoes a full eclipse must obey a pair of non-linear parametric equations which map any given eclipse geometry to a unique mass ratio, $q$, vs inclination relation\cite{Chanan1976}. Because we have already solved for a unique inclination given the primary and secondary eclipse geometry using the above expressions, this means we also obtain a unique mass ratio as a result of the Roche-lobe-filling assumption. We have also measured a velocity semi-amplitude of the donor, $K_2$, which is given by the binary mass function:

\begin{equation}
    \label{eq:BMF}
    \frac{M_\mathrm{WD} \sin^3(i)}{(1+q)^2}=\frac{P_\mathrm{b}\, K_{\mathrm{Donor}}^3}{2\pi \mathrm{G}}~,
\end{equation}
where G is the gravitational constant. Because we have already solved for a unique mass ratio, inclination, and have determined the orbital period, this expression yields a solution for $M_\mathrm{WD}$, which in combination with the mass ratio, also yields a solution for $M_\mathrm{Donor}$. Finally, since we have determined both component masses in the system, and know the orbital period, we can use Kepler's law to solve for the semi-major axis of the system:

\begin{equation}
    \label{eq:BMF}
    \frac{a^3}{P_\mathrm{b}^2}=\frac{G(M_{R_{\mathrm{WD}}}+M_{R_{\mathrm{Donor}}})}{4\pi^2}~,
\end{equation}
and thus, we use the semi-major axis in combination with the previously determined scaled radii $\frac{R_{\mathrm{WD}}}{a}$, $\frac{R_{\mathrm{Donor}}}{a}$ to compute the physical radii of the components, $R_{\mathrm{WD}}$ and $R_{\mathrm{Donor}}$. Thus, using the orbital period, four parameters measured from eclipse geometry, the Roche-lobe-filling nature of the donor and its radial velocity semi-amplitude, we are able to determine a unique solution for both component masses, radii, the semi-major axis of the system, the surface brightness ratio of the components and the orbital inclination.

When modelling the light curve, we used Gaussian priors constraining the radial velocity semi-amplitude based on the spectroscopic analysis. We sampled over the orbital inclination, $i$, the time of superior conjunction of the white dwarf, $T_0$, the mass of the white dwarf, $M_{\mathrm{WD}}$, the mass of the donor star, $M_{\mathrm{Donor}}$, the white dwarf to donor surface brightness ratio in each filter, $J$, and the radius of the white dwarf, $R_{\mathrm{WD}}$. We did not sample over the radius of the donor star, as we instead fixed it to fill the Roche lobe. We fixed the limb darkening and gravity darkening coefficients based on the temperature estimate of the white dwarf and donor star derived from atmospheric fits\cite{Claret2011,Claret2013,Claret2020,Claret2020b}.

\begin{figure}
\includegraphics[width=6.5in]{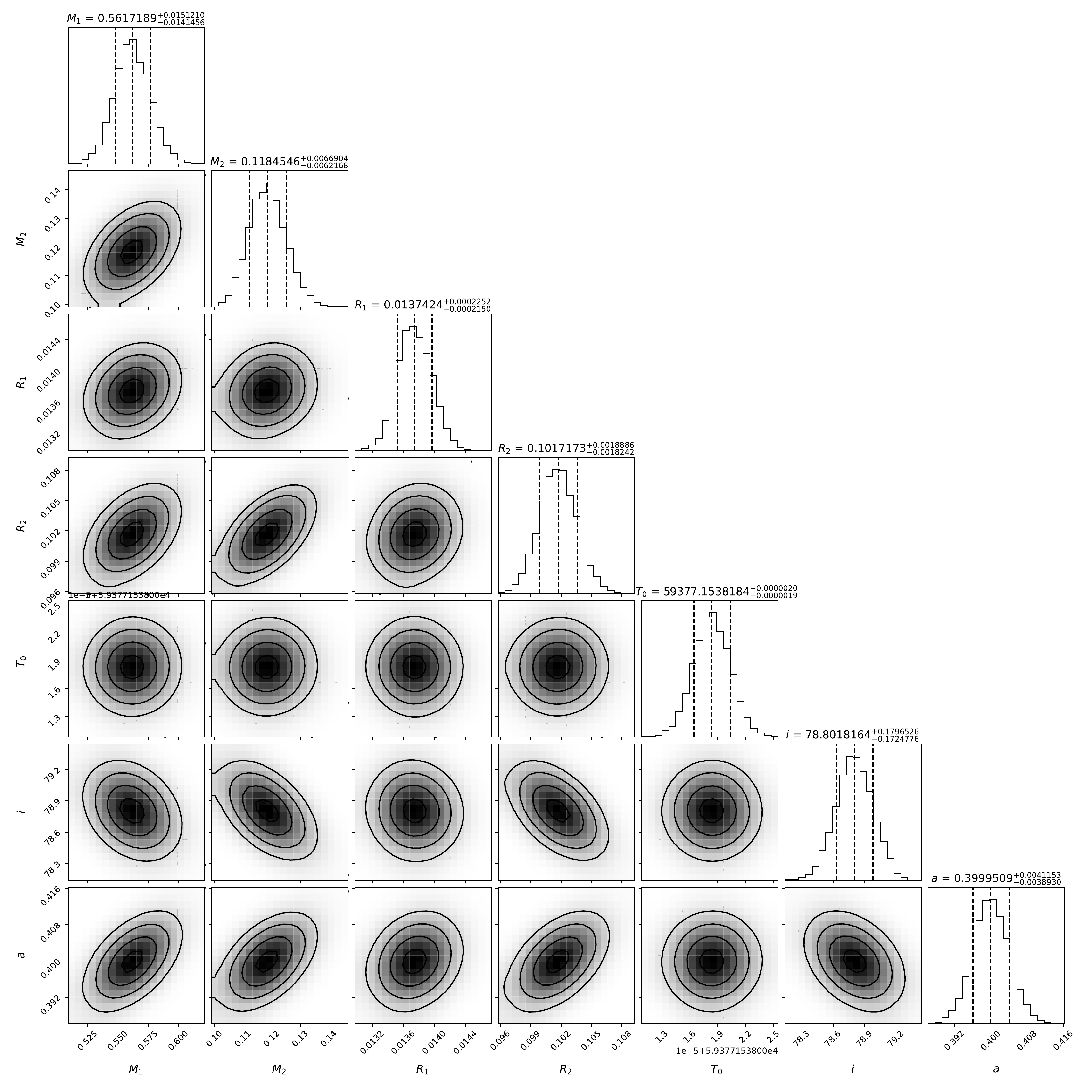}
\renewcommand{\figurename}{Extended Data Figure}
\linespread{1.3}\selectfont{}
\caption{A corner plot of some of the quantities derived from our final model, illustrating clean convergence in the distributions for all quantities. We would like to note that the radii are volume-averaged radii, and that $R_{2}$ was not directly sampled (instead, the two masses+inclination are sampled, and because the system is fully eclipsing $R_{2}$ is determined uniquely by the mass ratio and inclination). We also sampled over the surface brightness ratio J, but did not include it in these corner plots because we have five different Js (one for each filter–we avoided using a common J because doing this correctly requires atmospheric corrections in each passband, and the solutions for the other free params are largely independent of J). To ensure that using a different J for each filter was not influencing the other free params, we modeled all 5 filters independently, and found that they all converged to parameters in agreement with the combined fit.}
\label{fig:Corner}
\end{figure}

The strong O'Connell effect and contribution of the disk to the lightcurve makes it challenging to model the ellipsoidal variations in the system. We subtracted a term at the orbital frequency out of the light curve to account for this effect. However, we discovered there was a slight residual phase offset between the time of superior conjunction of the white dwarf inferred from these corrected ellipsoidal variations, and that indicated by the primary eclipse. 

Given that the system undergoes a full eclipse of the white dwarf and has a Roche-lobe-filling donor, rather than attempt to model the full light curve and account for the complexities in the ellipsoidal variations, we elected to model only points around the primary and secondary eclipses (within $\pm160\,\rm s$ of the mid-eclipse times), as the geometry of these contain all the information needed to fully constrain the system, and provide a more robust set of constraints than relying on ellipsoidal variations, which depend on quantities like the gravity darkening coefficients of the exotic donor star in each passband. The results of this modelling yielded the values reported in Table 1, and Figure \ref{fig:Eclipse_fit} illustrates the best-fit model overplotted with the primary and secondary eclipses, as well as the full lightcurve, while Figure \ref{fig:LC} illustrates the model fit to all the binned primary and secondary eclipses. All of our models included a first order linear polynomial with both a slope degree of freedom and arbitrary flux offset to minimize impacts of modulation due to the O'Connell effect near the eclipses on our models. We wish to emphasize that we did include ellipsoidal variations in our model, but also ran a test model with a second order polynomial fit around each eclipse to remove the sensitivity of our model to the ELVs (which can influence the inferred mass ratio), and other effects such as contributions from the accretion disk, effectively making the models only sensitive to the geometry of the eclipses of the donor and white dwarf themselves (since those contain all the information needed to constrain the system in a highly model-independent manner). We found that the parameter estimates derived from the models with the second order polynomial included were consistent with those derived by simply using a first order polynomial, indicating that effects such as the ELVs were not influencing the parameter estimates, and that the eclipses indeed are the dominate features influencing the constraints in the system.

\begin{figure}
\includegraphics[width=6.5in]{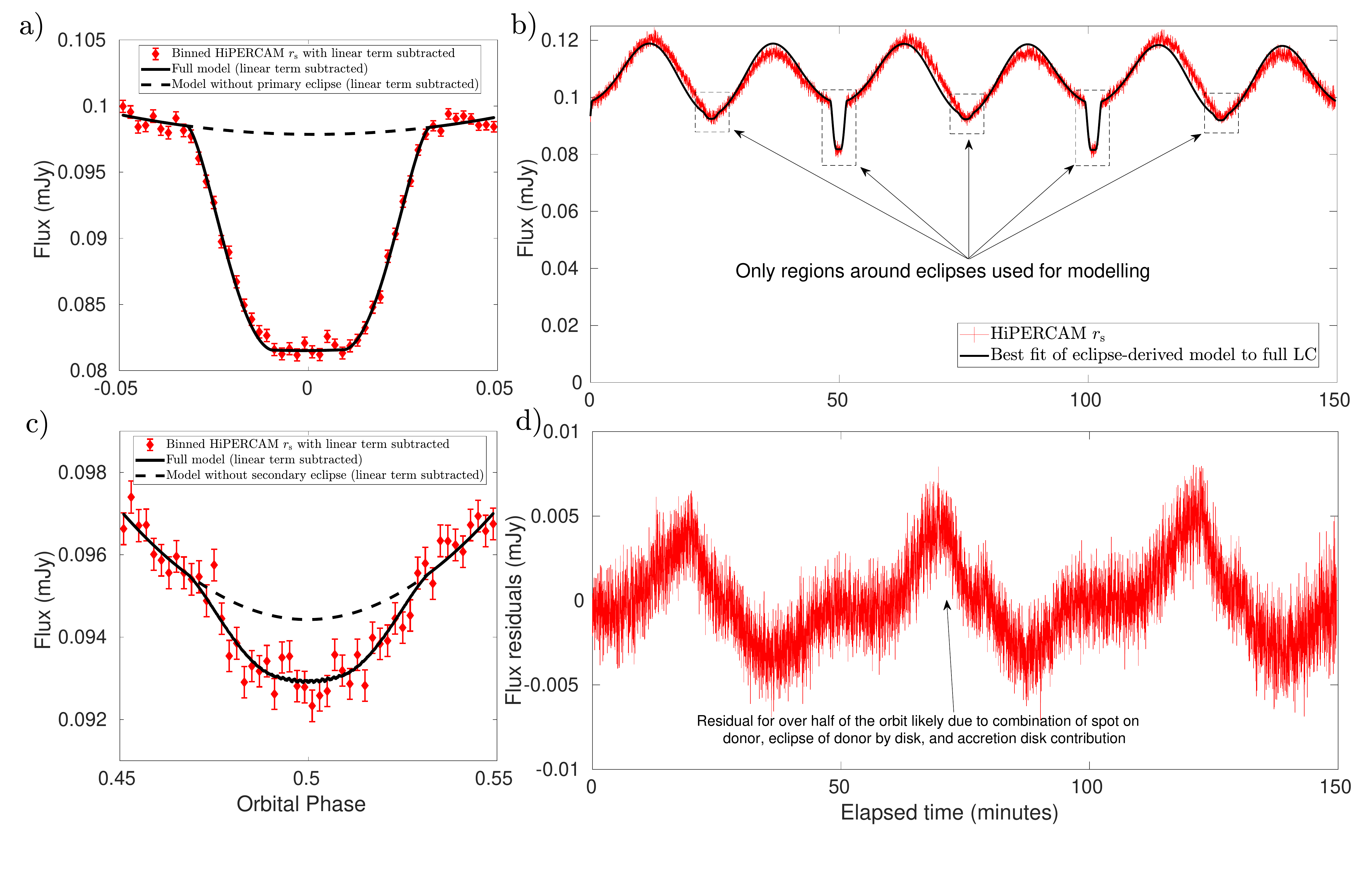}
\renewcommand{\figurename}{Extended Data Figure}
\linespread{1.3}\selectfont{}
\caption{\textbf{a)} Our best fit model of the primary eclipse, with the model shown as a solid black line, the binned data as red points, and the model without an eclipse as the dashed black line. In this figure, a linear polynomial component (of the functional form $y=a\times t$, where $t$ is the time from mid-eclipse (we applied separate corrections for the primary and secondary), has been subtracted out of both the model and the data for better visualization. We constructed the model by simultaneously fitting data around the primarily eclipse as well as data from the secondary eclipse (panel c). \textbf{b)} A best fit of the eclipse-derived model constructed from the eclipse data shown in panels a and c to the full HiPERCAM lightcurve. While the model roughly reproduces the correct amplitude of ellipsoidal variations, it does not to fully capture the structure seen in the full lightcurve. \textbf{c)} Our best fit model to the secondary eclipse, with a linear correction subtracted out of both the data and models. \textbf{d)} The residuals of the best fit of the eclipse-derived model to the full dataset shown in panel b. As is readily apparent, the strongest residuals occur out of eclipse, and likely arise due to a combination of effects from the accretion disk, and an O'Connell effect associated with the donor.}

\label{fig:Eclipse_fit}
\end{figure}

We used the Ultranest\cite{Buchner2014,Buchner2017,Buchner2021} package to perform the final parameter estimation by combining the light curve modelling of the eclipses and the donor's radial velocity semi-amplitude constraint. The derived parameters are reported in Table 1, and Extended Data Figure \ref{fig:Corner} illustrates the posterior distributions of these parameters.

As a sanity check regarding our assumptions, we conducted an independent modelling exercise, in which we did not assume a Roche-filling donor, but instead invoked a white dwarf mass radius relation\cite{Soares2017} for the accreting white dwarf, and we found that the models converged to values consistent with those derived from the models which assumed a Roche-filling donor. We expected this model to converge to similar values, because the values of the white dwarf mass and radius we found with our Roche-filling model agree remarkably well with a white dwarf mass-radius relation.

\subsection{Timing analysis}

Because of the substantial contribution of magnetic braking to the predicted orbital decay rate of the system, we undertook an effort to constrain the orbital decay rate, $\dot{P}_{\mathrm{b}}$, as we predict this value should be a factor of a few larger than what it would be due to purely general relativistic orbital decay, $\dot{P}_{\mathrm{b}}^{\mathrm{GW}}=(-4.27\pm0.24)\times 10^{-13}\rm \, s\,s^{-1}$. We use the HiPERCAM light curve in combination with archival data from ZTF, the Palomar Transient Factory (PTF) and the Catalina Real Time Transient Survey (CRTS) in order to constrain $\dot{P}_{\mathrm{b}}$. We find a value consistent with zero, with an uncertainty of $10^{-12}\rm \, s\,s^{-1}$, larger than the predicted value due to pure general relativity, indicating that we need to continue monitoring the system for the next several years to reach a degree of precision to test whether the orbital evolution is consistent with pure general relativity.

\subsection{Accretion disk analysis}

In order to estimate the white dwarf radial velocity semi-amplitude, we tried to obtain radial velocities from the centroid of the H$\alpha$ emission component of the accretion disk by fitting a pair of Lorentzians with positive amplitude and a symmetric splitting around a central radial velocity value. Because of the low SNR of the individual spectra and significant contamination from the H$\alpha$ absorption line of the donor star (and possibly from the white dwarf itself), we were unable to extract a radial velocity for the accretor in this manner. However, we used the masses estimated from our model to construct a coadded spectrum in the rest frame of the accretor, as seen in Figure \ref{fig:Red_Spectrum}, and then fit the accretion disk profile to determine the splitting between the red and blue components in the disk. We find a splitting of $17.33\pm0.49\rm\,\AA$, which corresponds to a velocity of $810\pm23\rm \, km\,s^{-1}$. Assuming a Keplerian disk, this corresponds to a radius of $0.163\pm0.009\,R_{\rm \odot}$ around the white dwarf. This is approximately 79 percent the radius of the the white dwarf's Roche lobe, a regime in which material should not be able to maintain a stable orbit\cite{Paczyinski1971}. We verify that we find a similar result with the He I line at $6678\,\rm\AA$. It is possible that the splitting in the disk lines is underestimated due to underlying contamination from the donor and WD absorption lines, but our results suggest that the matter producing the emission in this system exists near the limit of where a stable accretion disk can form in the white dwarf's Roche lobe.

We also investigated the object's ZTF power spectrum to look for signatures of a peak associated with a superhump excess, which could allow for an independent measurement of the mass ratio in the system. However, the only peaks we were able to identify in the power spectrum were those associated with the orbital period and beat frequencies of the orbital period and the sidereal day.

Curiously, the accretion disk does not manifest itself in a deep eclipse of the donor, which is something we would expect for an optically thick accretion disk. This may suggest that the instantaneous mass transfer rate of the system is lower than the longer term average mass transfer rate predicted by our MESA models, as some models would predict an optically thick accretion disk given the estimated average mass transfer rate. This is in contrast to some other sub-period minimum systems such as GALEX J194419.33+491257.0\cite{Kato2014b}, which exhibits signatures of a higher mass transfer rate, though we emphasize that the donor in ZTF J1813+4251 is significantly more evolved than any other known sub-period minimum CV, and thus its donor properties may also be influencing its mass-transfer behavior in a unique manner.

\subsection{Kinematic Analysis}

We performed a kinematic analysis of ZTF J1813+4251's orbit in the Galaxy (plotted in Extended Data Figure \ref{fig:height}) and find that it is consistent with residing in the Galactic thick disk. This suggests that the progenitor stars of the system formed several billion years ago and likely had a low metallicity (though we emphasize that we do not believe the low metallicity value inferred from spectroscopic fitting of the donor is reliable given its highly exotic current state). We used the galpy\cite{Bovy2015} package to compute its trajectory around the Milky Way over 6 Gyr, using the McMillan2017 potential\cite{McMillan2017}.

\begin{figure}
\includegraphics[width=6.0in,trim={0 8cm 1cm 3cm},clip]{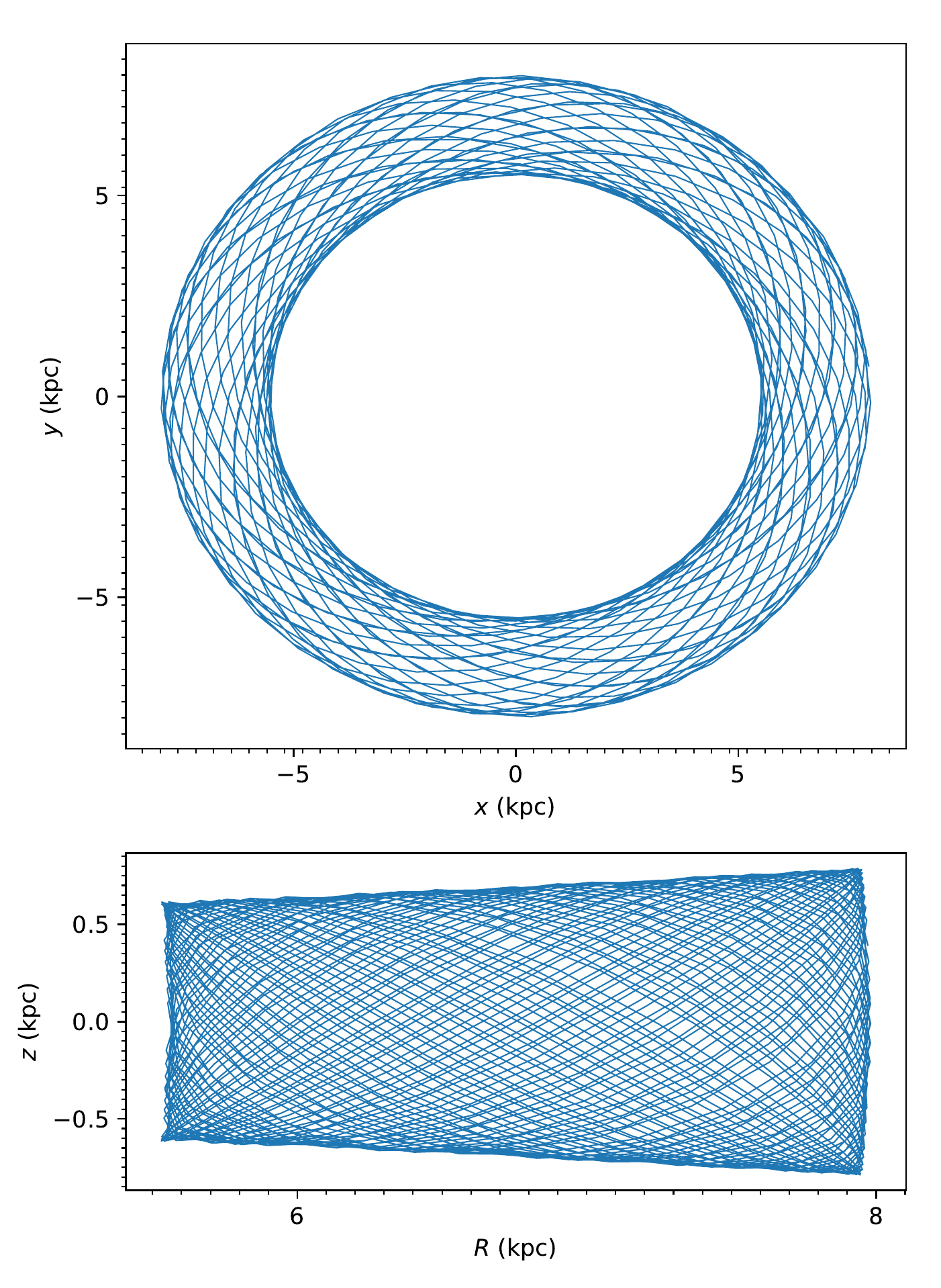}
\renewcommand{\figurename}{Extended Data Figure}
\linespread{1.3}\selectfont{}
\caption{A set of panels illustrating the orbit of ZTF J1813+4251 around the Milky Way. The system is consistent with residing in the Galactic thick disk, orbiting between 5 and 8 kpc from the Galactic center, within half a kpc of the Galactic disk in height.}

\label{fig:height}
\end{figure}

\subsection{MESA models}
\label{sec:mesa_details}

To investigate the formation history and future evolution of ZTF J1813+4251, we calculated a grid of binary evolution models using MESA (version r12778) \cite{Paxton2011, Paxton2015}. These models build on those calculated by \cite{El-Badry2021}, who used them to interpret observations of evolved CVs with longer periods (2-6 hours). See also the extensive grid of CV models, including ultra short period systems with He donors, generated  with MESA in Kalomeni et al. (2016)\cite{Kalomeni2016}.

We initialized the models with a detached, circular-orbit binary containing a $1.1\,M_{\odot}$ zero-age main sequence star (the donor) and a $0.55\,M_{\odot}$ point mass representing the white dwarf. Following \cite{Wong2021}, we use the Skye + CMS equation of state.
Orbital angular momentum is removed via gravitational wave radiation and magnetic braking.  Roche lobe radii are computed using the fit from \cite{Eggleton1983}, and mass transfer rates during Roche lobe overflow are determined following the prescription from \cite{Kolb1990}. We model mass transfer as being fully non-conservative, with the mass that is transferred from the donor to the white dwarf eventually being lost from the vicinity of the latter. The orbital separation evolves under the assumption that the mass that is lost has the same specific angular momentum as the accretor. Although mass transfer in real CVs is not {\it instantaneously} non-conservative, the mass accreted by the white dwarf is expected to be ejected from the system by classical nova explosions, which recur on a timescale that is short compared to the timescale on which the orbit evolves.

Magnetic braking follows the prescription from \cite{Rappaport1983}, with $\gamma_{\rm mb}=3$. Following \cite{Podsiadlowski2002}, we assume that the magnetic braking torque is exponentially suppressed when the donor's convective envelope contains less than 2~percent of the star's mass. This is the default implementation of magnetic braking in MESA as of version r15140; it effectively makes magnetic braking inefficient when the donor's effective temperature is $T_{\rm eff} \gtrsim 6000\,\rm K$. We turn magnetic braking off at the period minimum. 

We explored models with initial periods ranging from 0.5 to 3 day. At fixed donor mass, longer initial periods lead to donors that are older, and hence more evolved, at the onset of mass transfer. In Extended Data Table~\ref{tab:mesa} and Figure~\ref{fig:mesa} we highlight models which have similar properties to  ZTF J1813+4251 at a period of 51 minutes. These models begin mass transfer just as the donor is completing its main sequence evolution. Models with shorter initial periods begin mass transfer before the donor has undergone significant nuclear evolution; these evolve into normal CVs. Models with significantly longer initial periods begin mass transfer as the donor is evolving up the giant branch and terminate their evolution as extremely low-mass white dwarfs with long periods (in the case of stable Roche lobe overfilling) or short periods (in the case of common envelope evolution).

In all the models shown in Figure~\ref{fig:mesa} there is an initial period of thermal-timescale mass transfer in which the mass transfer rate reaches $\sim 10^{-7}\,M_{\odot}\,\rm yr^{-1}$ and most of the donor's outer envelope is lost. By the time the models reach a period of 6 hours, all that remains of the donor is a helium core and a puffy hydrogen envelope containing a few $\times 10^{-2}\,M_{\odot}$ of hydrogen. This envelope is steadily removed as the orbit shrinks, still driven primarily by magnetic braking. At periods of a few hours, the donors in these models are very similar to the evolved CVs in the \cite{El-Badry2021} sample. They heat up as their envelope is removed, with the more evolved models (those in which mass transfer begins later) being hotter at fixed orbital period. 

By the time the models reach $P_{\rm b} = 51$ minutes, their surface is predicted to be $\sim 75$~percent helium by mass. The surface nitrogen-to-carbon ratio is predicted to be $\sim 2,000$ times the solar value, because the present-day surface of the donor was previously inside the CNO-burning core while the donor was on the main sequence. The predicted mass transfer rates and accretor temperatures at 51 minutes differ significantly between the three models we show: the more evolved models with higher $T_{\rm eff}$ have less efficient magnetic braking and thus lower mass transfer rates. 

We calculate the effective temperature of the accretor assuming it is set by compressional heating, as described by Equation 2 of \cite{Townsley2009}. This makes the accretor temperature a sensitive probe of the time-averaged accretion rate. Both the donor and accretor temperatures of  ZTF J1813+4251 are best-matched by the model with $t_{\rm RLOF}/t_{\rm MS}=0.95$. Here $t_{\rm MS}$ is the main-sequence lifetime the donor would have if it were an isolated star, and $t_{\rm RLOF}$ is the age of the system at the onset of Roche lobe overflow. In this model, magnetic braking is still the dominant mode of angular momentum loss, leading to a predicted orbital inspiral that is a factor of $\sim 3$ larger than expected from gravitational radiation alone. 

The models reach a minimum period between 13 and 22 minutes. After this, their orbits begin to widen and they transition to primarily helium mass transfer with fully-degenerate donors. The predicted inspiral time is $\approx$ 70 Myr for the model with $t_{\rm RLOF}/t_{\rm MS} = 0.95$. Observational constraints on the period derivative will test these models, allowing us to directly measure the relative importance of magnetic braking and gravitational waves in removing angular momentum from the system. 

\begin{figure*}
    \centering
    \includegraphics[width=0.9\textwidth]{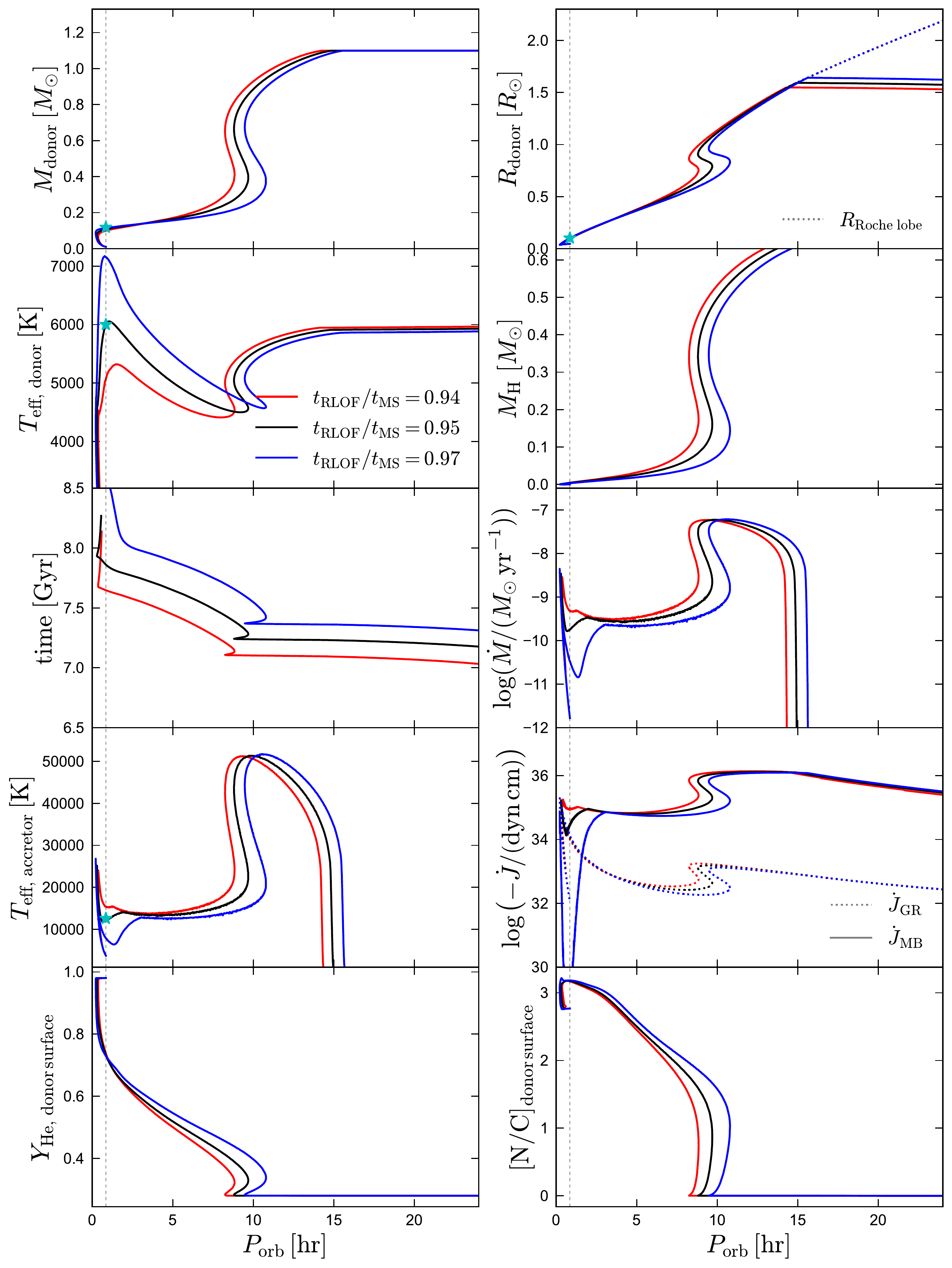}
    \renewcommand{\figurename}{Extended Data Figure}
    \caption{MESA binary evolution models (Extended Data Table~\ref{tab:mesa}). Red, black and blue lines show models that overflow their Roche lobes after 94, 95 and 97 percent of the donor's main-sequence lifetime. The dashed vertical line shows 51 minutes.  }
    \label{fig:mesa}
\end{figure*}

\captionsetup[table]{name=Extended Data Table}

\begin{table}
\setcounter{table}{0}  
\caption{MESA models. All models have an initial donor mass of $1.1\,M_{\odot}$ and a white dwarf mass of $0.55\,M_{\odot}$. The three models correspond to the red, black and blue lines in Extended Data Figure~\ref{fig:mesa}. }
\label{tab:mesa}
\begin{tabular}{llllllll}
$P_\mathrm{b}^{\rm init}$ & $t_{\rm RLOF}/t_{\rm MS}$ & $P_\mathrm{b}^{\rm RLOF}$ & $T_{\rm eff,\,Donor}$ & $M_{\rm Donor}$ & $P_\mathrm{b}^{\rm min}$   & $\log(\dot{M}/M_{\odot}\,\rm yr^{-1})$          & $\rm Time\,\,since\,\,RLOF$  \\ 
\hline
$[\rm day]$   &                           & $[\rm day]$   & $[\rm K]$   & $M_{\odot}$          & $[\rm min]$ &  & $[\rm Gyr]$ \\
\hline
\hline
$2.63$           & $0.94$                      & $0.60$           & $5068$        &  $0.100$        & $22$              & $-9.31$                    & $0.55$                         \\
$2.69$           & $0.95$                      & $0.63 $          & $6009$        &  $0.107$       & $18$              & $-9.75$                    & $0.63$                         \\
$2.75$           & $0.97$                      & $0.66$           & $7156$        &  $0.115$        & $13$              & $-10.46$                   & $1.26$                        
\end{tabular}
\end{table}

\end{methods}

\section{Data Availability}
Reduced HiPERCAM photometric data, LRIS spectroscopic data, and MESA tracks resulting from the models are available at \\ https://github.com/kburdge/ZTFJ1813-4251.git. The ZTF used data are all in the public domain. The proprietary period for the spectroscopic data will expire at the start of 2022, at which point the raw spectroscopic images will also be accessible via the Keck observatory archive. 

\section{Code Availability}

Upon request, the first author will provide code (primarily in Python) used to analyze the observations, create MESA models, and any data used to generate figures (MATLAB was used to generate most of the figures). The LCURVE modelling code can be found at: https://github.com/trmrsh/cpp-lcurve




\end{document}